\documentclass{elsarticle}
\usepackage{etex}
\usepackage{amssymb}
\usepackage{amsmath}
\usepackage{mathrsfs}
\usepackage{stmaryrd}
\usepackage{latexsym}
\usepackage{hhline}
\usepackage{enumerate}
\usepackage[latin1]{inputenc}
\usepackage{calc}
\usepackage{graphicx}
\usepackage{ifthen}
\usepackage{pst-all}
\usepackage{pst-poly}
\usepackage{multido}
\usepackage[normalem]{ulem}
\usepackage{tikz}
\usepackage{hyperref}

\usepackage{todonotes} 

\newif\iflongversion
\longversionfalse

\newtheorem{proposition}{Proposition}
\newtheorem{lemma}{Lemma}
\newtheorem{theorem}{Theorem}
\newtheorem{corollary}{Corollary}
\newdefinition{remark}{Remark}
\newtheorem{claim}{Claim}
\newproof{proof}{Proof}

\newenvironment{subproof}
{ 
  \newcommand{\qedsymbol}{$\blacksquare$}
  \begin{proof}
}{
  \end{proof}
}

\newdefinition{definition}{Definition}
\newdefinition{example}{Example}

\begin{document}
\title{Non-erasing Chomsky-Sch{\"u}tzenberger theorem with grammar-independent alphabet\tnoteref{t1}\tnoteref{t2}}
\tnotetext[t1]{This research was partially supported by PRIN Project 2010LYA9RH and  by CNR-IEIIT.}
\tnotetext[t2]{Part of this work was published  preliminarily in~\cite{DBLP:conf/lata/Crespi-Reghizzi16} and presented at the \emph{Conf. dedicated to the scientific legacy of M.P. Sch{\"u}tzenberger} \cite{ScLegacyMPSchutz:Crespi-Reghizzi16}.}

\author[pdm,cnr]{Stefano {Crespi Reghizzi}}
\ead{stefano.crespireghizzi@polimi.it}
\author[pdm,cnr]{Pierluigi {San Pietro}}
\ead{pierluigi.sanpietro@polimi.it}

\address[pdm]{Dipartimento di Elettronica Informazione e Bioingegneria, Politecnico di Milano,\\Piazza Leonardo da Vinci 32, Milano, Italy}
\address[cnr]{CNR IEIIT-MI, Milano, Italy}

\begin{abstract}
The famous theorem  by Chomsky and Sch{\"u}tzenberger (CST) says that every context-free language 
$L$ over an alphabet $\Sigma$ is representable as $h(D \cap R)$, where $D$ is a Dyck language over a set $\Omega$ of brackets, $R$ is a  local 
language and $h$ is an alphabetic homomorphism that erases unboundedly many symbols.
Berstel found that the number of erasures can be linearly limited if the grammar is in Greibach normal form; 
Berstel and Boasson (and later,  independently, Okhotin) proved a non-erasing variant of CST for grammars in Double Greibach Normal Form. 
In  all these CST statements, however, the size of the Dyck alphabet $\Omega$ depends on the grammar size for $L$. 
In the  Stanley variant of the CST, $|\Omega|$ only depends on $|\Sigma|$ and not on the grammar, 
but the homomorphism erases many more symbols than in the other versions of CST; also, the regular language $R$ is strictly locally testable but not local. 
We prove a new version of CST which combines both features of being non-erasing and of using a grammar-independent alphabet. 
In our construction, $|\Omega|$ is polynomial in $|\Sigma|$, namely $O(|\Sigma|^{46})$, and the regular language $R$ is strictly locally testable.
Using a recent generalization of Medvedev's homomorphic characterization of regular languages, we prove that the degree in the polynomial dependence of $|\Omega|$ on $|\Sigma|$ 
may be reduced to just 2 in the case of linear grammars in Double Greibach Normal Form. 
\end{abstract}

\maketitle

\section{Introduction}
In formal language theory,  the  idea of  homomorphic characterization of  a language family refers 
to the definition of all and only the  languages in that family,  starting from languages of simpler families and applying an alphabetic transformation.  
Such idea has been applied to many different language families, from the regular to the recursively enumerable ones, 
and also to non-textual languages, such as the two-dimensional picture languages~\cite{GiammRestivo1997}.
Our focus here is on context-free (\emph{CF}) languages, 
but a short reference to the earlier homomorphic characterization of  regular languages, known as Medvedev theorem~\cite{Medvedev1964} (also in~\cite{Eilenberg74}), is useful to set the ground. 
The  regular language family coincides with the family obtained by applying an alphabetic letter-to-letter homomorphism to the simpler family  
at that time named  \emph{definite events} and presently known as {\em local}, and also referred to as \emph{strictly locally testable} languages of width 2, shortened to 2-\emph{SLT} (e.g.,~\cite{Caron2000}). 
\par
Then, Chomsky and Sch{\"u}tzenberger ~\cite{ChomskySchutz1963} stated the theorem, referred to as \emph{CST},  
saying that the CF family coincides with the family obtained by the following two steps. First, we intersect a Dyck language $D$ over an alphabet consisting of brackets, and a 2-SLT language $R$. Second, we apply  to the result an alphabetic homomorphism $h$, in formula $h(D \cap R)$, which maps some brackets to terminal letters and erases some others. 
Therefore, a word $w\in D \cap R$ may be longer than its image $h(w)$. 
\par
The original proof of CST considers a grammar in Chomsky Normal Form (CNF) and uses a Dyck  alphabet made 
by a distinct pair of  brackets for each grammar rule, which makes the Dyck alphabet  typically much larger than the terminal alphabet and dependent on the grammar size.
\par
In the almost contemporary variant by Stanley~\cite{Stanley1965} (also in  Ginsburg~\cite{Ginsburg1966}), the Dyck alphabet is grammar-independent: it consists of the terminal alphabet, a marked copy thereof, 
and four extra letters, two of them used as delimiters (i.e., brackets), the other two as unary codes.  
In this variant, the homomorphism has to erase many more symbols than in the original version of CST. 
The regular language is not 2-SLT, but it is immediate to prove that it is  strictly locally testable, 
by using a width parameter greater than two, depending linearly on the  number of grammar rules. 
\par
Then Berstel~\cite{Berstel79} (his Exercise 3.8) found that fewer symbols than in the original CST need to be erased by the homomorphism, if the grammar is in Greibach Normal Form.  In that case, there exists a constant $k>0$ such that, for every word $w\in D \cap R$, the ratio of the lengths of $w$ and $h(w)$ does not exceed $k$. Later, Berstel and Boasson~\cite{DBLP:journals/fuin/BerstelB96}, and  independently  Okhotin~\cite{Okhotin2012}, proved a non-erasing variant of CST by using  grammars in
Double Greibach Normal Form, \emph{DGNF} (see e.g.~\cite{Engelfriet1992}).
\par
In the statements ~\cite{Berstel79},~\cite{DBLP:journals/fuin/BerstelB96} and~\cite{Okhotin2012}, 
however, the Dyck alphabet depends on the grammar size.
Most formal language books  include   statements and proofs of CST essentially similar to  the early ones.
\par
To sum up, we may classify  the existing versions of CST with respect to two primary parameters: 
the property of being erasing  versus nonerasing, and the grammar-dependence versus grammar-independence of the Dyck alphabet, as shown in the following table:
\begin{center}
\begin{tabular}{|l||p{4.1cm}|p{3.5cm}|}\hline
& \multicolumn{2}{|c|}{Dyck Alphabet}\\ \hline\hline
  Homomorphism & \emph{grammar-dependent} & \emph{grammar-independent } \\\hline
 \emph{erasing }	&  Chomsky and Sch{\"u}tzenberger~\cite{ChomskySchutz1963}, & Stanley~\cite{Stanley1965}\\
 &Berstel~\cite{Berstel79}&\\
\hline
 \emph{nonerasing}	& Berstel and Boasson~\cite{DBLP:journals/fuin/BerstelB96}, Okhotin~\cite{Okhotin2012} &
\\\hline
\end{tabular}
\end{center}
This paper fills the empty case of the table. It  presents a new non-erasing version of CST that uses a Dyck alphabet polynomially dependent on the terminal alphabet size  and independent from the grammar size. Besides the two parameters of the table, a third aspect may be considered: whether the regular language is strictly locally testable or not and, in the former case, whether its width is two or greater. Actually, this aspect is correlated with the alphabet choice, because, if the alphabet is grammar-independent, the  grammar complexity, which cannot be encoded inside the Dyck alphabet, must affect the size of the regular language, in particular its SLT width. 
We show that the width parameter is logarithmically related to grammar complexity, both in the erasing and the non-erasing cases.
\par
In our previous communication~\cite{DBLP:conf/lata/Crespi-Reghizzi16}  we proved  by means of standard constructions for pushdown automata, grammars and sequential transductions 
(without any optimization effort) that the Dyck alphabet needed by our version of CST is polynomially related to the original alphabet. However, we could not give a precise upper bound.
Here we develop some new grammar transformations (in particular a new normal form that we call \emph{quotiented}) and analyze their complexity, to obtain a precise, but still pessimistic, upper bound on the exponent of the polynomial dependence between the two alphabets.  
As a side result, we improve  the known transformation from CNF to the \emph{generalized} DGNF~\cite{DBLP:journals/ipl/Yoshinaka09,DBLP:journals/tcs/BlattnerG82}) in the relevant case here, namely when the two parameters of the DGNF are equal, i.e., when the terminal prefix and suffix of every production right-hand side have the same length.
\par
 The  Dyck alphabet we use, though independent from the grammar size, is much larger than the original alphabet.  At the end, we show that a substantial reduction of alphabet size is easy  in  the case of the linear grammars in DGNF. For that we  exploit  the recent extension ~\cite{DBLP:journals/ijfcs/Crespi-ReghizziP12} of Medvedev homomorphic characterization of  regular languages, which reduces the alphabet size at the expense of the SLT width. 
\par
The enduring popularity of  CST  can be ascribed to the elegant combination of two structural aspects of CF languages, namely  the free well-nesting of brackets, and a simple finite-state control on the adjacency of brackets. 
Taking inspiration from CST, many homomorphic characterizations for other language families have been proposed. A commented historical bibliography is in \cite{ScLegacyMPSchutz:Crespi-Reghizzi16}; we mention one example, the case of the slender CF languages \cite{DBLP:journals/actaC/DomosiO01}.
\par
Paper organization: Sect.~\ref{SectionPreliminaries} lists the basic definitions, recalls some relevant CST formulations,  and proves a trade-off between the Dyck alphabet size and the regular language size.
Sect.~\ref{SectHomCharSuitableLengthLang} proves CST using a grammar-independent alphabet and a non-erasing homomorphism; it first introduces and studies the size of the grammar normal forms needed, then it develops the main proof, and at last presents an example.  Sect.~\ref{SectHomCharBasedOnMedvedev} states the Dyck alphabet size for CF grammars in the general case, and shows that a much smaller Dyck alphabet suffices for the linear CF grammars in DGNF.  The  conclusion mentions directions for further research.

\section{Preliminaries and basic properties}\label{SectionPreliminaries}
For brevity, we omit  most classical definitions (for which we refer primarily to ~\cite{Harrison1978} and ~\cite{Berstel79}) and just list our notation. 
Let $\Sigma$ denote a finite terminal alphabet and $\varepsilon$ the  empty word. For a word $x$, $|x|$ denotes the length of $x$;
the $i$-th letter of $x$ is  $x(i)$, $1\le i\le |x|$, i.e., $x = x(1)x(2) \dots x(|x|)$. 
For every integer $r>0$, the language $\Sigma^{<r}$ is defined as $\{x \in \Sigma^* \mid |x| < r\}$, and similarly for $\Sigma^{\leq r}$ and $\Sigma^r$.
Notice that $|\Sigma^{<r}| \in O(|\Sigma|^r)$. The reversal of a word $x$ is denoted by $x^R= x(|x|)\dots x(2)x(1) $. The right quotient of a language $L\subseteq \Sigma^*$ by a word $w \in \Sigma^*$ is denoted by $L_{/w}= \{x \mid xw \in L \}$.
\par
For finite alphabets $\Delta, \Gamma$, an \emph{alphabetic homomorphism} is a mapping $h: \Delta \to \Gamma^*$; if, for some $d\in \Delta$, $h(d)=\varepsilon$,  then $h$ is called \emph{erasing}, while it is called  strict or \emph{letter-to-letter} if, for every $d\in \Delta$, $h(d)$ is in $\Gamma$.
\par 
Given a nondeterministic finite automaton (NFA) $A$,  the language recognized by $A$ is denoted by $L(A)$.
The \emph{size} of a regular language $R=L(A)$, $\textit{size}(R)$,  is the number of states of a minimal NFA that recognizes the language.
\par
A \emph{context-free} (CF) \emph{grammar} is a 4-tuple $G=(\Sigma, N, P, S)$ where $N$ is the nonterminal alphabet, $P\subseteq N \times (\Sigma\cup N)^*$ is the rule set, and $S\in N$ is the axiom.
Since we only deal with context-free grammars and languages, we often drop the word ``context-free''. For simplicity, in this paper we define the size of $G$ to be the number $|N|$ of the nonterminals of $G$.
The language generated by $G$ starting from a nonterminal $X \in N$ is $L(G,X)$;  we shorten $L(G,S)$ into $L(G)$. 
A word in $L(G)$ is also called a \emph{sentence}.
\par
A  grammar is \emph{linear} if the right side of each rule contains at most one nonterminal symbol.
\par
A  grammar is in \emph{Chomsky normal form} (CNF) if the right side of each rule is in $\Sigma$ or in $NN$.
\par
A grammar $G=(\Sigma, N, P, S)$ is in {\em Double Greibach normal form} (DGNF) if the right side of each rule is in $\Sigma$ or in $\Sigma N^* \Sigma$. 
It is in {\em cubic} DGNF if it is in DGNF and there are at most three nonterminals in the right side of every rule.
A generalization of DGNF is the $(m,n)$-GNF (see, e.g., 
\cite{DBLP:journals/ipl/Yoshinaka09,DBLP:journals/tcs/BlattnerG82}) where the right-hand side of each rule is in $\Sigma^m N^* \Sigma^n$ or in $\Sigma^{< m+n}$, for $m,n\ge 1$.
\par
The family SLT  of {\em strictly locally testable} languages
\cite{McPa71} is next defined,
dealing only with $\varepsilon$-free languages.
For every word $w\in \Sigma^+$, for all $k\ge 2$, let $i_k(w)$ and
$t_k(w)$ denote the prefix and, resp., the suffix of $w$ of
length $k$ if $|w|\ge k$, or $w$ itself if $|w|<k$. For $k\ge |w|$, let $f_k(w)$
denote the set of factors of $w$ of length $k$.
Extend $i_k, t_k, f_k$ to languages as usual.
\begin{definition}\label{defk-SLT}
Let $k\geq 2$. A language $L$ is $k$-{\em strictly locally testable} ($k$-\emph{SLT}),
if there exist finite sets
 $W \subseteq\Sigma \cup \Sigma^2 \cup \dots \cup \Sigma ^{k-1}$,
$I_{k-1},T_{k-1}\subseteq \Sigma ^{k-1}$, and
$F_{k}\subseteq \Sigma ^{k}$ such that for every $x\in \Sigma^+$,
$x\in L$ if, and only if,
\[
x \in W \;\lor \;
\big(i_{k-1}(x)\in I_{k-1}\, \wedge \,
t_{k-1}(x)\in T_{k-1} \,\wedge\,
f_{k}(x)\subseteq F_{k}\big).
\]
A language is {\em strictly locally testable (SLT)} if it is $k$-SLT for
some $k$, called its {\em width}.
\end{definition}
Value $k=2$ yields the  well-known family of {\em local
languages}. The SLT family  is
strictly included in the family  of \emph{regular languages} and forms a hierarchy with respect to the width. The size of a $k$-SLT language over $\Sigma$ is in $O(|\Sigma|^k)$.

\subsection{Past statements of CST}\label{subsubsectPastCSTs}
The following notation for Dyck alphabets and languages is from~\cite{Okhotin2012}.
For any finite set $X$, the \emph{Dyck alphabet} is the set, denoted by $\Omega_X$,
of brackets labeled with elements of $X$:
\[
\Omega_X = \left\{ \, [_x \, \mid x \in X \right\} \,\cup\, \left\{ \, ]_x \, \mid x \in X \right\}.
\]
The \emph{Dyck language} $D_X \subset{} \Omega^*_X$ is generated by the following grammar:
\begin{equation}
S \to [_x \, S\, ]_x \text{ for each } x \in X , \quad S \to SS, \quad S \to \varepsilon
\label{eqDyckGra}
\end{equation}
The notation $\Omega_X$ for the Dyck alphabet should not be confused with the asymptotic lower bound notation $\varOmega$, which is also used later in the paper.
\par
Let $k= |X|$. Clearly, each Dyck language $D_X$ is isomorphic to $D_{ \{1,\dots,k\}}$. For brevity we write $\Omega_k$ and $D_k$ instead of $\Omega_{\{1,\dots,k\}}$ and $D_{ \{1,\dots,k\} }$, respectively.
\par
Since it is obviously impossible for an odd-length sentence to be the image of a Dyck sentence under a letter-to-letter homomorphism,
the CST variant by  Okhotin  (Th. 3 in~\cite{Okhotin2012}) modifies the Dyck language by adding \emph{neutral} symbols to its alphabet, and we do the same here.
\begin{definition}\label{defAlphabetDyckWithNeutral}
Let $q, \l \geq 1$. We denote  by $\Omega_{q,l}$ an alphabet containing $q$ pairs of brackets and $l$ distinct symbols, called \emph{neutral}~\cite{Okhotin2012}.
The \emph{Dyck language with neutral symbols} over alphabet $\Omega_{q,l}$, denoted by $D_{q,l}$, 
is the language generated by the  grammar in Eq. \eqref{eqDyckGra}, enriched with the rules $S \to c$, for each neutral symbol $c$ in $\Omega_{q,l}$.
\end{definition}
\par
We need two of the known statements of CST, the non-erasing version by Berstel and Boasson~\cite{DBLP:journals/fuin/BerstelB96}, which we present following Okhotin~\cite{Okhotin2012}, and the fixed alphabet version by Stanley~\cite{Stanley1965}: they are respectively reproduced as Th.~\ref{th-1-Okhotin} and Th.~\ref{ThStanleyNostro}. 

Moreover,  we  prove in Th.~\ref{th-okhotinNostro} a simple statement about the exact number of brackets needed in Okhotin's construction and a slight generalization of his theorem, which will be useful in later proofs.
\begin{theorem}\label{th-1-Okhotin}
(Th. 1 of Okhotin~\cite{Okhotin2012})
	A language $L\subseteq \left(\Sigma^2  \right)^*$ is context-free if, and only if, there exist an integer $k>0$, a regular language $R\subseteq \Omega_k^*$  and  a letter-to-letter  homomorphism $h : \Omega_k \to \Sigma$  such that $L = h\left( D_k \cap R \right)$.  
\end{theorem}
\par
Following Okhotin's Lemma 1~\cite{Okhotin2012}, since $L\subseteq \left(\Sigma^2  \right)^*$, we can assume that a grammar for $L$ is in  \emph{even}-DGNF, that is the grammar form such that the right side of each rule is in $\Sigma N^* \Sigma$.

\begin{theorem}\label{th-okhotinNostro}(Derived from the proof of Th. 1 of ~\cite{Okhotin2012})
\begin{enumerate}\item  Let  $L\subseteq \left(\Sigma^2  \right)^*$  be the language  defined by a grammar $G=(\Sigma, N, P, S)$  in  even-DGNF, and let $k=  |P|^2+|P|$. Then, there exist a regular language $R\subseteq \Omega_k^*$  and  a letter-to-letter  homomorphism $h : \Omega_k \to \Sigma$  such that $L = h\left( D_k \cap R \right)$.  
\item Let  $G=(\Sigma, N, P, S)$  be an even-DGNF grammar and let $q = |P|^2+|N|\cdot |P|$. Then, there exists a letter-to-letter  homomorphism $h : \Omega_q \to \Sigma$  such that, 
 for all $X \in N$, there is a regular language $R_X$ satisfying the  equality
$L(G,X)  = h\left( D_q \cap R_X \right)$. 
\end{enumerate}
\end{theorem}
\begin{proof}
To prove part (1), we revisit the proof in~\cite{Okhotin2012}. 
We assume that $L$ is generated by a CNF grammar and we convert it into an even-DGNF grammar $G=(\Sigma,N,P,S)$. 
 Each  rule has thus the form $A \to b C_1 \ldots C_n d$, where one can further assume that nonterminals $C_1, \dots, C_n$ are pairwise distinct. 
The leftmost terminal $b$ and the rightmost terminal $d$ in this rule is replaced, respectively, by an open  or a closed  bracket.
Each bracket is labeled with a pair of rules of $P$ of the form 
$\langle X \to \xi_1 A \xi_2, A \to b C_1 \ldots C_n d\rangle$, 
the first component being the ``previous'' rule $X \to \xi_1 A \xi_2$ (where, as assumed, $\xi_1 \xi_2$ have no occurrence of $A$), and the second one the ``current'' rule 
$A \to b C_1 \ldots C_n d$ itself. The idea is to represent the derivation step $X  \Longrightarrow \xi_1 A \xi_2 \Longrightarrow \xi_1 b C_1 \ldots C_n d \xi_2$.

Since there is no rule deriving the axiom of the grammar, we need also a distinguished label, which can just be the axiom itself, in the first component of the leftmost open bracket, e.g.,
the label of the first opening bracket can be a pair of the form $\langle S, S \to a B_1 \dots B_n c\rangle$.

Therefore, the value $k$ of the Dyck alphabet $\Omega_k$ is at most $|P|^2+|P|$, i.e., in $O\left( |P|^2\right)$.

Incidentally, an example of a Dyck sentence corresponding to a derivation is shown in Sect. \ref{sect:example}, Eq. \eqref{eqExampleOkhotinHomo}, while 
Eq.~\eqref{eq-ex-h-hom} illustrates the definition of homomorphism $h$.
\par
Part (2) can be proved, by applying, for every $X \in N$, the preceding proof to the grammar, denoted by  $G_X= (\Sigma, N, P, X)$, which is obtained from $G$ by selecting $X$ as the axiom.
It follows that there exist an integer $k=|P|^2+|P|$, a letter-to-letter homomorphism $h_X: \Omega_k\to \Sigma$, and a regular language $R_X\subseteq \Omega_k^+$, such that $L(G,X) = h_X(D_k\cap R_X)$.
The Dyck alphabet of  $L(G,X)$ is thus composed of $|P|^2 + |P|$ pairs of brackets; however, for every $X$, each language $L(G,X)$ is defined by essentially the same grammar, except for the axiom $X$. 
Therefore, the Dyck language for $L(G,X)$, is defined by the same set of $|P|^2$ bracket pairs, each labeled with a pair of productions of $P$, already defined  for the Dyck set of $L(G)$; and by
$|P|$ bracket pairs  whose first component is labeled $X$.

We can now define the Dyck alphabet $\Omega_q$ as the union of all the above Dyck alphabets; therefore,  the total number of bracket pairs is $q = |P|^2+|N|\cdot |P|$.
\par\noindent Notice the language $R_X$ differs from $R_Y$, for $X\neq Y$, only in that it must start and end with a bracket whose label has as first component $X$ rather than $Y$.
\par\noindent 
 At last, it is immediate to define one 
letter-to-letter homomorphism $h$ which is valid for the CST of each language $L(G,X)$.
\qed
\end{proof}
\par
Furthermore,  the regular language $R$ produced in Okhotin's proof  is a 2-SLT language:  $R$ simply checks that every pair of adjacent brackets corresponds to the correct consecutive application of two  rules in a leftmost derivation, and that the leftmost (open) bracket and the rightmost (closed) bracket are labeled with the axiom.
\par
Next we state Stanley's CST, as presented in~\cite{Ginsburg1966}, and add an immediate consequence.
\begin{theorem}\label{ThStanleyNostro}(derived from Th. 3.7.1 of Ginsburg~\cite{Ginsburg1966})
Given an alphabet $\Sigma$, there exist a Dyck alphabet $\Omega$ and an alphabetic erasing homomorphism $h : \Omega^* \to \Sigma^*$  which satisfy the following properties:
\begin{enumerate}
	\item for each language $L \subseteq \Sigma^*$, $L$ is context-free if, and only if, there exists a regular language $R \subseteq \Omega^*$  such that $L= h(D \cap R)$;
	\item if  $L=L(G)$, with $G=(\Sigma, N, P, S)$ in CNF, then there exists a constant $k$ with $k \in O(|P|)$ such that $R$ is a $k$-SLT language.
\end{enumerate}
\end{theorem}
\begin{proof}
We only need to prove item (2), which is not considered in~\cite{Ginsburg1966}.   The Dyck alphabet in~\cite{Ginsburg1966} is 
$
\Omega = \Sigma\, \cup\, \Sigma' \, \cup \, \{c, c',d, d'\}
$, 
where $\Sigma'$ is a primed copy of $\Sigma$; thus $|\Omega| = 2 |\Sigma| + 4$. Homomorphism $h$ erases  any letter in $\{c, c',d, d'\} \cup \Sigma'$ 
and maps the other letters on the corresponding terminal letter. Ginsburg lists a right-linear grammar for $R$ that has rules of the following types:
\begin{equation}
\begin{array}{l}
X \to aa', \text{ if } X\to a \in P
\\
 X\to aa' d' c'^i d' B , \; \text{ if } X \to a \in P\; \text{ and }i\text{ is the label of a rule } E\to AB 
\\
E \to d c^i d A, \;\text{ if } i\text{ is the label of a rule } E\to AB
\end{array}
\label{eqGinsburghRLgramm}
\end{equation}
Notice that $dc^id$ and $d'c'^id'$, $1< i < |P|$,  represent the integer $i$, i.e., a grammar rule label, as a unary code. 
Clearly, any sentence generated by the right-linear grammar satisfies a locally testable constraint, namely that any two adjacent codes  are compatible with the above rules. 
Since the code length is at most  $|P|+ 2$, a sliding window of width $2|P|+ 2$ suffices to test the constraint.
\qed
\end{proof}
\par
A new  straightforward improvement on Stanley theorem can be  obtained  using a binary code instead of a unary one, to represent grammar rule labels in base two.
This allows for a sliding window of size logarithmic in the number of productions.
\begin{corollary}\label{CorollaryStanley}	
Under the assumptions in Th.~\ref{ThStanleyNostro},
there exists a constant $k$, with $k \in \mathcal{O}(\log|P|)$, such that $R$ is a $k$-SLT language.
\end{corollary}
\begin{proof}
 A sketch of the proof suffices. Suppose that the original grammar has $n>0$ rules, and let $h = \lceil \log (n)\rceil$, 
where $\log$ is the base 2 logarithm, and assume that symbols $0,1$ are not in  $\Sigma$. 
Given $i$, $0\le i\le n$,
let $\llbracket i \rrbracket_2$ be the  representation of number $i$ in base two using $h$ bits, which is a word over alphabet $\{0,1\}$.  
Modify  grammar \eqref{eqGinsburghRLgramm} for the regular language $R$ as follows:  replace every rule of the form 
$ X\to aa' d' c'^i d' B$ with  $X\to aa' d' \llbracket i \rrbracket_2 d' B$, and  replace every rule of the form  $E \to d c^i d A $
with $E \to d \llbracket i \rrbracket^R_2 d A$, where $ \llbracket i \rrbracket^R_2$ is the mirror image of  $\llbracket i \rrbracket_2$. 
By taking $k=2h+2$ it is immediate to see that the regular language defined by this grammar is $k$-SLT.\qed
\end{proof}
Encoding   grammar rules by positional numbers is also the key idea applied in Sect.~\ref{SectHomCharSuitableLengthLang}, but, since the homomorphism is not allowed to erase such  numbers, 
a  much more sophisticated  representation will be needed. 

\subsection{Trade-off between Dyck alphabet and regular language sizes}
It is worth contrasting the two  versions of CST reproduced as Th.~\ref{th-1-Okhotin} and Cor.~\ref{CorollaryStanley}:  the former uses a larger Dyck alphabet and a simpler regular language, while the latter has a smaller Dyck alphabet and a more complex regular language. With a little thought, it is possible to formulate a precise   relation of general validity between the Dyck alphabet size, the complexity of the regular language, and the number of nonterminal symbols of the CF grammar. 
\par
We recall the language family $\{M^{(m)}\}$, $m>0$, defined for each $m$ as the language: 
\begin{equation}
M^{(m)} = (ab)^* \cup (aab)^* \cup \dots \cup (a^nb)^* 
\nonumber
\end{equation}
By a classical result of Gruska~\cite{Gruska67}, every CF grammar generating $M^{(m)}$ must have at least $m$ nonterminal symbols.
Although $M^{(m)}$ is regular, it is easy to transform it into a non-regular CF language $L^{(m)}$ having the same property, e.g., 
 $L^{(m)} = \{ w w^R \mid w \in M^{(m)} \}$. It is obvious that  every grammar for $L^{(m)}$ needs at least $m$ nonterminal symbols. 
\par
The following proposition gives a lower bound on the size of the Dyck alphabet and on the size of the minimal NFA accepting $R$.

\begin{proposition}\label{propos:TradeOff}
For every finite alphabet $\Sigma$ with $|\Sigma|>1$, for every $m>0$ there exists a language $L\subseteq \Sigma^*$ such that 
every  context-free grammar for $L$ has at least $m$ nonterminals and, 
for every  homomorphic characterization as 
$L=h\left(D\cap R\right)$ (with $D, R \subseteq \Omega^*$ for some Dyck alphabet $\Omega$),  the following relation holds:
\[
\left| \Omega \right| \cdot \textit{size}^2(R) \ge m .
\]
\end{proposition}
\proof
It suffices to outline the proof. 
For every $m>0$, let $\Omega^{(m)},\, h^{(m)}:\Omega^{(m)}\to\Sigma ,\, R^{(m)}$ be, respectively, a Dyck alphabet,  a  homomorphism and a regular language such that:
$L^{(m)} = h^{(m)}\left(D^{(m)}\cap R^{(m)}\right)$, 
where $L^{(m)}$ is a CF language whose grammar requires at least $m$ nonterminal symbols.
\par\noindent
First, we construct a grammar $G$ for language  $D^{(m)}\cap R^{(m)}$ by means of the classical construction in \cite{BarHillel61} (Th. 8.1), 
which assumes that 
each right part of a rule in the grammar is either a terminal character or a nonterminal word. A straightforward grammar in this form 
for the Dyck language $D^{(m)}$ has exactly $\left|\Omega^{(m)}\right|+1$ nonterminals. 
Then, the number of nonterminals of grammar $G$ is at most $\left|\Omega^{(m)}\right|\cdot \textit{size}^2(R)$. 
At last, by a straightforward transformation of $G$,  we 
obtain a grammar $G^{(m)}$ defining language $h(L(G))=L^{(m)}$ and having the same number of nonterminals as $G$. \qed

%
\par
It is worth observing that, if a CST characterization of $L$ is such that  the alphabet size $|\Omega|$  depends on the alphabet size $|\Sigma|$ but  does not depend on the number $m$ of nonterminals, 
then it follows that
a minimal NFA for the regular language $R$ must  have a number of states dependent on the number of nonterminals, i.e., $R$ must reflect the size of a grammar for $L$. 
In this case, a simple asymptotic lower bound on the number of states of a NFA for $R$ is clearly $\varOmega(\sqrt{m})$, 
i.e., the square root of the number of nonterminals of a minimal grammar for $L$. 
Obviously, in general this lower bound may be too small and $\textit{size}(R)$ may actually be quite larger: for instance, in the case of  Th.~\ref{ThStanleyNostro}, the regular language $R$ has an NFA  recognizer  with $O(m^2)$ states.

\section{New homomorphic characterization}\label{SectHomCharSuitableLengthLang}
The section starts with the grammar normal forms to be later used in the proof of the CST, and examines their descriptive complexity, with the aim of obtaining at least an estimation of the size of the Dyck alphabet (which  in~\cite{DBLP:conf/lata/Crespi-Reghizzi16} was just proved to be polynomial in the terminal alphabet size). 
Then  the section continues with the main theorem and its proof, and terminates with an example illustrating the central idea of the proof.

\subsection{Preliminaries on grammar normal forms}\label{subsectNormalForms}
We revisit the classic construction of DGNF starting from a CNF grammar, to establish a numerical relation between the size of the two grammars. Then we introduce a new normal form, called quotiented, and combine it with the DGNF form.
\par
The following lemma supplements a well-known result about DGNF (e.g.,~\cite{AutebertBerstelBoasson1997}) with an explicit upper bound,  which is lacking in the literature, on the size of the equivalent DGNF grammar in terms of the original CNF grammar size.

\begin{lemma}\label{lm-DGNF}
Given a CNF grammar $G=(\Delta, N,P,S)$, over  a finite alphabet $\Delta$, there exists an equivalent grammar $\widetilde{G}=(\Delta, \widetilde{N},\widetilde{P},\widetilde{S})$ in cubic DGNF (thesis of Theor. 3.4 of~\cite{AutebertBerstelBoasson1997}). 
Grammar $\widetilde{G}$ is such that 
$\left|\widetilde{N}\right| \in O\left(\left|\Delta\right| \cdot \left|N\right|^2\right)$ and 
$\left|\widetilde{P}\right| \in O\left(|\Delta|^6\cdot |N|^8  \right) $.
\end{lemma}
\begin{proof}
Starting from the  construction of $\widetilde{G}$ in~\cite{AutebertBerstelBoasson1997}, we estimate its size.
The construction involves four steps, but we only need the first three, since the last step computes a quadratic form  not needed here. 
Leftmost and rightmost derivations are respectively denoted by $\Rightarrow_L$ and $\Rightarrow_R$. We compute the sizes as we proceed.
\par
\textbf{Step 1} defines $\widetilde{N}$ as the union of $N$ and the finite set $\mathcal{H}$ next defined.
\begin{gather*}
\forall a\in\Delta, X\in N, \text{ let } L(a,X)= \left\{m \in N^* \mid X \stackrel * \Rightarrow_L \alpha \Rightarrow a m  \text{ where } \alpha \in N^*\right\};
\\
\forall a\in\Delta, X\in N, \text{ let }  R(X,a)= \left\{m \in N^* \mid X \stackrel * \Rightarrow_R \alpha \Rightarrow  m a \text{ where } \alpha \in N^*\right\};
\\
\mathcal{L}= \left\{L(a,X)  \mid a\in\Delta, X\in N \right\} 
		 \text{ hence }\left|\mathcal{L}\right| \leq |\Delta| \cdot |N| ;
\\
\mathcal{R}= \left\{R(X,a)  \mid a\in\Delta, X\in N \right\}
		 \text{ hence }\left|\mathcal{R}\right| \leq |\Delta| \cdot |N| ;
\\
\text{ hence }\left|\mathcal{L}\cup \mathcal{R}\right| \leq 2|\Delta| \cdot |N|;
\\
\mathcal{H} = \text{ closure of }\mathcal{L}\cup \mathcal{R} \text{ under the right and left quotients by a letter of }N ;
\\
\text{ it follows that } \left|\mathcal{H}\right| \leq 4 |\Delta| \cdot |N|^2. 
\end{gather*}
Therefore, it holds:
\begin{equation}
\left|\widetilde{N}\right| \in O\left(|\Delta| \cdot |N|^2\right).
\label{eq:SizeWidetildeN}
\end{equation}
\par
Then \textbf{Step 2} and \textbf{Step 3} construct a cubic DNGF grammar $\widetilde{G}$, over the terminal alphabet $\Delta$ and the nonterminal alphabet $\widetilde{N}$, 
i.e., the rules $\widetilde{P}$ are in $\widetilde{N} \times \Delta \widetilde{N}^{\le 3}\Delta$.
A rough and quick calculation, obtained from Eq. \eqref{eq:SizeWidetildeN} supposing that all nonterminals can be combined in all ways, yields the (pessimistic) estimation:
\[
\left|\widetilde{P}\right| \in O\left(|\Delta|^2\left|\widetilde{N}\right|^4\right) = 
O\left(|\Delta|^2\cdot |\Delta|^4  |N|^8  \right) =
O\left(|\Delta|^6\cdot |N|^8  \right).
\]
\qed
\end{proof}

We now want to generalize Lm.~\ref{lm-DGNF} to an $(m,m)$-GNF, for values of $m$ larger than 3, since this form is convenient for proving our CST. Unfortunately, the grammar transformation algorithms known to us are not adequate here (as next explained), and we have to introduce 
a new normal form for grammars, called \emph{quotiented} and then to prove an intermediate lemma. 
\par
We observe that, exploiting 
known results on GNF (see, e.g.,~\cite{DBLP:journals/ipl/Yoshinaka09,DBLP:journals/tcs/BlattnerG82}), it is fairly obvious that every CNF grammar can be transformed into an $(m,n)$-GNF whose size is polynomially related to the size of the original grammar.
For instance, the very simple construction provided in~\cite{DBLP:journals/ipl/Yoshinaka09} shows that, if 
$G=(\Sigma, N, P, S)$ is in CNF, then an equivalent $(m,n)$-GNF grammar can be built such that the nonterminal alphabet is in $O(|N|^2)$ and the number of rules is in $O\left(|\Sigma|^{m+n+2} \cdot  |N|^{2m+2n+4}\right)$.
Unfortunately, although the latter relation is polynomial in the size of the grammar,  both terms, featuring the base $|\Sigma|$ or the base $|N|$, exhibit an undesirable exponential dependence on $m+n$.
\par
In contrast, anticipating our Lm. \ref{lm-partitionedCNF}, under the assumption  $m=n$, i.e.,  for  $(m,m)$-GNF grammars, we obtain that the number of rules is in $|\Sigma|^{O(m)}\cdot O(|N|^8)$, 
i.e.,  
the term with base $|\Sigma|$ has still an exponential dependence in the value $m$, but the the term with base $|N|$ has instead a polynomial dependence.

While for the term with base $|\Sigma|$ the preceding exponential dependence of the number of rules in the value $m$ remains, the exponential dependence  disappears for the term with base $|N|$,
more precisely, the number of rules is in $|\Sigma|^{O(m)}\cdot O(|N|^8)$.
\par
We  define the new normal form.
\begin{definition}\label{def-quotientedNF}
A grammar $G=(\Sigma, N, P, S)$ is  \textbf{quotiented of order} $r\ge 1$ if the axiom $S$ does not occur in any right-hand side and the set $P$ of rules is partitioned in two sets, $P_q, P_r$, such that:
\[
P_q \subseteq \{S\} \times N \cdot \Sigma^{<r} 
\quad \quad \text{ and }\quad  P_r \subseteq \left(N-\{S\} \right)\times \left(N\;\cup\; \Sigma^r \right)^*.
\] 

The two sets  include, respectively, the rules for the axiom, and the rules for the other nonterminals.  

If $G$ is quotiented of order $r$, then it is said to be, for the same order $r$ :
\begin{description}
	\item[quotiented CNF ] (\emph{Q-CNF}),  if 
$P_r\subseteq \left(N \times N^2 \right)$
\; $\cup\; \left( N\times \{\Sigma^r \cup\varepsilon\}\right)$;
	\item[quotiented DGNF ](\emph{Q-DGNF}),  if 
$P_r\subseteq \left(N \times \Sigma^r N^*\Sigma^r\right)$
\, $\cup\, \left( N\times \{\Sigma^r \cup\varepsilon\}\right)$.
\end{description} 
\end{definition}

\begin{example}
We show three equivalent quotiented forms with $r=3$.
\[
\renewcommand{\arraystretch}{1.2}
\begin{tabular}{p{2cm}|c |c }
																& $P_r$ &  $P_q$
\\\hline
quotiented  & $X \to aaa X bbb X \mid \varepsilon,\quad 
																											Y \to abb Y \mid abb$  & 
\\\cline{1-2}
Q-CNF  &  $X \to X_1 X_2,\,  X_1\to X_3 X,\, X_2\to X_4 X \mid \varepsilon$ &
\\ &$X_3 \to aaa, \, X_4 \to bbb$  &   $S \to X aa \mid Yb$
\\ & $Y \to X_5 Y \mid abb, \,  X_5 \to abb$ & 
\\\cline{1-2}
Q-DGNF & $X \to a^3 XZX b^3\mid a^3  b^3\mid \epsilon$,  $Z \to    b^3 a^3$ & 
\\& $Y \to abb Y abb \mid abb \mid \varepsilon $ &
\end{tabular}
\]
\end{example}
For a quotiented grammar, if the rule $S \to X w$, where $X \in N$ and $w\in \Sigma^{<r}$, is in $P$, then the language  generated starting from $X$ is included in $L(G)_{/ w}$, which is the right quotient of $L(G)$ by $w$.  
\par
The next lemma studies the complexity of  the  Q-CNF and Q-DGNF normal forms. Since its proof operates on an  alphabet made by tuples of letters, we need  the following definition.
\begin{definition}[Tuple alphabet and homomorphism]\label{DefTupleAlphabet}
For an  alphabet $\Sigma$,  let $\Delta_r=  \left\{ \langle a_1, \dots, a_r\rangle \mid a_1, \dots, a_r \in \Sigma \right\}$ for all $r\geq 2$.
An element of the alphabet $\Delta_r$ is called an $r$-\emph{tuple} or simply
a tuple. 
\par\noindent 
The \emph{tuple homomorphism} $\pi_{r}: \Delta_{r} \to \Sigma^+$ is defined by 
\[
\pi_{r}\left( \langle a_1, \dots, a_r \rangle \right) = a_1 \dots a_r,\;\text{ for } a_1, \dots, a_r\in \Sigma.
\]
\end{definition}
\par\noindent
The inverse morphism $\pi^{-1}_{r}$  transforms a language included in  $\left(\Sigma^r\right)^+$  into a language of $r$-tuples; it 
 will be applied for constructing an $(r,r)$-GNF grammar.
\par
Historical remark. In our earlier paper~\cite{DBLP:conf/lata/Crespi-Reghizzi16} we already proved that, for every CNF grammar, there exists an  equivalent Q-CNF grammar $G'$. The proof applied  standard transformations back and forth from grammars to pushdown automata,  and  a suitable finite-state transduction.
That approach has two drawbacks: first, the resulting complexity of the  grammar,
although polynomial in $|N|\cdot |\Sigma|^{r}$, is very high and difficult to compute with precision. 
Second,  the  overly general constructions employed in that proof barred any significant improvement in the complexity.
To overcame  such limitations, we present a new direct construction of the Q-CNF and of the Q-DGNF grammars, which allows us to prove the better (but still very pessimistic) upper bounds of  
Eq. \eqref{L4}, \eqref{L5} and  \eqref{L6}, 
and may open the way for further improvements.

\begin{lemma}\label{lm-partitionedCNF}
For every grammar $G=(\Sigma, N,P,S)$ in CNF, for every $r\ge 1$, there exist an equivalent Q-CNF grammar  $G'=(\Sigma, N', P', S')$ 
and an equivalent Q-DGNF grammar $G''=(\Sigma, N'', P'', S'')$, both of order $r$, such that:
\begin{gather}
\label{L4} |N'| \in O( |N|\cdot |\Sigma|^{2r}) 
\\
\label{L5} |P'|  \in O(|P|\cdot |\Sigma|^{3r});
\\
\label{L6} |P''|  \in O\left(|\Sigma|^{6r}\cdot |N'|^8\right) = O\left(|\Sigma|^{22r}\cdot |N|^8\right).
\end{gather}
\end{lemma}
\begin{proof}
\noindent{\em Construction of $G'$.}  Let $\dashv$ be a new symbol not in $\Sigma$. 
The  set $N'$ of nonterminal symbols is composed of $S'$, 
and of the set of  3-tuples:
\[
N \times \Sigma^{<r}\times \Sigma^{<r}\quad \cup\quad  N \times \Sigma^{<r}\times \Sigma^{<r}\dashv.
\]
The tuples  have the following intuitive meaning: a nonterminal of $N'$ of the form $\langle A, u, \, w\rangle$  generates a word that entirely stays inside  a word of $L(G)_{/w}$, while
 a nonterminal of the form $\langle A, u, \, w\dashv\rangle$ generates a word that protrudes into a suffix $w\in \Sigma^{<r}$. 
\par
Thus, $|N'|$ is in $O(|N|\cdot |\Sigma^{<r}|^2)$, i.e., since  $|\Sigma^{<r}|\in O(|\Sigma|^r)$, it holds: 
\begin{equation}\label{N'}
|N'| \in  O(|N|\cdot |\Sigma|^{2r})
\end{equation}

\par  
The grammar rules are next defined.
First, for every $w \in \Sigma^{<r}$, the rule $S' \to \langle S,\epsilon,w\dashv\rangle\,  w $ is in $P'$ if $L(G)_{/w} \neq \emptyset $.
\\
Second, the remaining rules of $P'$ are defined, for all $A, B, C \in N\setminus \{S\}$, for all $a\in \Sigma$, for all $t,u,v,w \in \Sigma^{<r}$,  by the following clauses: 
\begin{gather}
\nonumber
\langle S,\epsilon,w\dashv\rangle  \to 
\langle A, \varepsilon, \, t\rangle 
\langle B, t, \, w\dashv  \rangle
\text{ if } S\to AB \in P
\\
\nonumber
\langle A, u, \, v\rangle \to 
\langle B, u, \, t \rangle
\langle C, t , v\rangle
\text{ if } A\to BC \in P
\\ 
\nonumber
\langle A, u, \, w\dashv\rangle \to 
\langle B, u, \, t \rangle
\langle C, t , w\dashv\rangle
\text{ if } A\to BC \in P
\\
\nonumber
\langle A, \varepsilon, \, w\dashv\rangle \to \varepsilon 
\\ \label{R2}
\langle A, u, \, \varepsilon \rangle \to  ua
\text{ if } A\to a \in P \text{ and } |ua| = r
\\ \label{R3}
\langle A, u,ua\,  \rangle \to \varepsilon 
\text{ if } A\to a \in P \text{ and } |ua|<r
\end{gather}
We assume that any rule containing  unreachable and undefined nonterminals is removed from $P'$.
\par
 For all  $A\in N$,
for all  $x, y  \in \left(\Sigma^r\right)^* $,  for all $t,u,v\in\Sigma^{<r}$, and for $z$ such that  $uz \in \Sigma^{<r}$,  we claim that 
grammar $G'$ has the derivation  
\begin{equation}
S' \Longrightarrow  \langle S,\epsilon,w\dashv\rangle  w 
\stackrel * \Longrightarrow  x\,  \,\langle A, u, v \rangle \, v  y  w 
\stackrel * \Longrightarrow  x\, uz \, v y w
 \label{eq:quotientedDerivation}
\end{equation}
if, and only if, grammar $G$ has the derivation 
\begin{equation}
 S \stackrel * \Longrightarrow_G x\, u A y\, w 
\stackrel * \Longrightarrow_G x\, u \,z\,v\,y\,w 
 \label{eq:originalDerivation}
\end{equation}
The two derivations are schematized in Fig.~\ref{figSchemaDerivazioniQuozientate}.
\par
\begin{figure}[tbh!]
\begin{center}
\begin{tikzpicture}[xscale=2.8 ]
\draw[|-|] (0,0)-- (0.4,0);
\draw[|-|] (0.4,0)--(0.8,0);
\draw[dotted] (0.8,0)--(1.2,0);
\draw[|-|] (1.2,0)--(1.6,0);
\draw[dotted] (1.6,0)--(2.4,0);
\draw[|-|] (2.4,0)--(2.8,0);
\draw[|-|,dotted] (2.8,0)--(3.2,0);
\draw[|-|] (3.2,0)--(3.6,0);
\draw[|-|] (3.6,0)--(4.0,0);
\draw[|-|] (1.6,0.1)-- (1.8,0.1);
\draw[<->] (1.8,-0.8)--(2.25,-0.8);
\draw[|-|] (2.25,0.1)-- (2.4,0.1);
\draw[<->] (0,-0.8)--(1.6,-0.8);
\draw[<->] (2.4,-0.8)--(4.0,-0.8);
\draw[|-|] (4.0,0)-- (4.3,0);

\draw[thick] (0.8,-0.3)node[below] {$x\in \left(\Sigma^r\right)^*$};
\draw[thick] (1.7,0.3)node {$u$};
\draw[thick] (2.04,-0.3)node[below] {$z$};
\draw[thick] (2.3,0.3)node {$v$};
\draw[thick] (3.4,-0.3)node[below] {$y\in \left(\Sigma^r\right)^*$};
\draw[thick] (4.25,0.3)node {$w\in\Sigma^{<r}$};

\draw[-] (2.1,2.0)-- (1.8,0.1);\draw[-] (2.1,2.0)-- (2.4,0.1);
\draw[thick] (2.1,2.2)node[above] {$A$};
\draw[-] (2.1,3.3)-- (0.0,0.1);\draw[-] (2.1,3.3)-- (4.3,0.1);
\draw[thick] (2.1,3.3)node[above] {$S$};

\end{tikzpicture} 
\vspace{10mm}
\begin{tikzpicture}[xscale=2.8 ]
\draw[|-|] (0,0)-- (0.4,0);
\draw[|-|] (0.4,0)--(0.8,0);
\draw[dotted] (0.8,0)--(1.2,0);
\draw[|-|] (1.2,0)--(1.6,0);
\draw[dotted] (1.6,0)--(2.4,0);
\draw[|-|] (2.4,0)--(2.8,0);
\draw[|-|,dotted] (2.8,0)--(3.2,0);
\draw[|-|] (3.2,0)--(3.6,0);
\draw[|-|] (3.6,0)--(4.0,0);
\draw[|-|] (1.6,0.1)-- (1.8,0.1);
\draw[|-|] (2.25,0.1)-- (2.4,0.1);
\draw[<->] (0,-0.8)--(1.6,-0.8);
\draw[<->] (1.8,-0.8)--(2.25,-0.8);
\draw[<->] (2.4,-0.8)--(4.0,-0.8);
\draw[|-|] (4.0,0)-- (4.3,0);

\draw[thick] (0.8,-0.3)node[below] {$x\in \left(\Sigma^r\right)^*$};
\draw[thick] (1.7,0.3)node {$u$};
\draw[thick] (2.04,-0.3)node[below] {$z$};
\draw[thick] (2.3,0.3)node {$v$};
\draw[thick] (3.4,-0.3)node[below] {$y\in \left(\Sigma^r\right)^*$};
\draw[thick] (4.25,0.3)node {$w\in\Sigma^{<r}$};

\draw[-] (2.1,2.0)-- (1.6,0.1);\draw[-] (2.1,2.0)-- (2.25,0.1);
\draw[thick] (2.1,2.2)node[above] {$\langle A, u, v \rangle$};
\draw[-] (2.1,3.6)-- (0.0,0.1);
\draw[-] (2.1,3.6)-- (4.0,0.1);
\draw[thick] (2.1,3.6)node[above] {$\langle S,\epsilon,w\dashv\rangle $};

\draw[-] (2.1,5.0)-- (2.1,3.9);\draw[-] (2.1,5.0)-- (4.3,0.1);
\draw[thick] (2.1,5.0)node[above] {$S'$};
\end{tikzpicture}
\end{center}
\caption{Scheme of the original grammar derivation (top) and the corresponding quotiented grammar derivation (bottom), respectively described in Eq. \eqref{eq:originalDerivation} and Eq. \eqref{eq:quotientedDerivation}  of the proof of Lm. \ref{lm-partitionedCNF}.}\label{figSchemaDerivazioniQuozientate}
\end{figure}
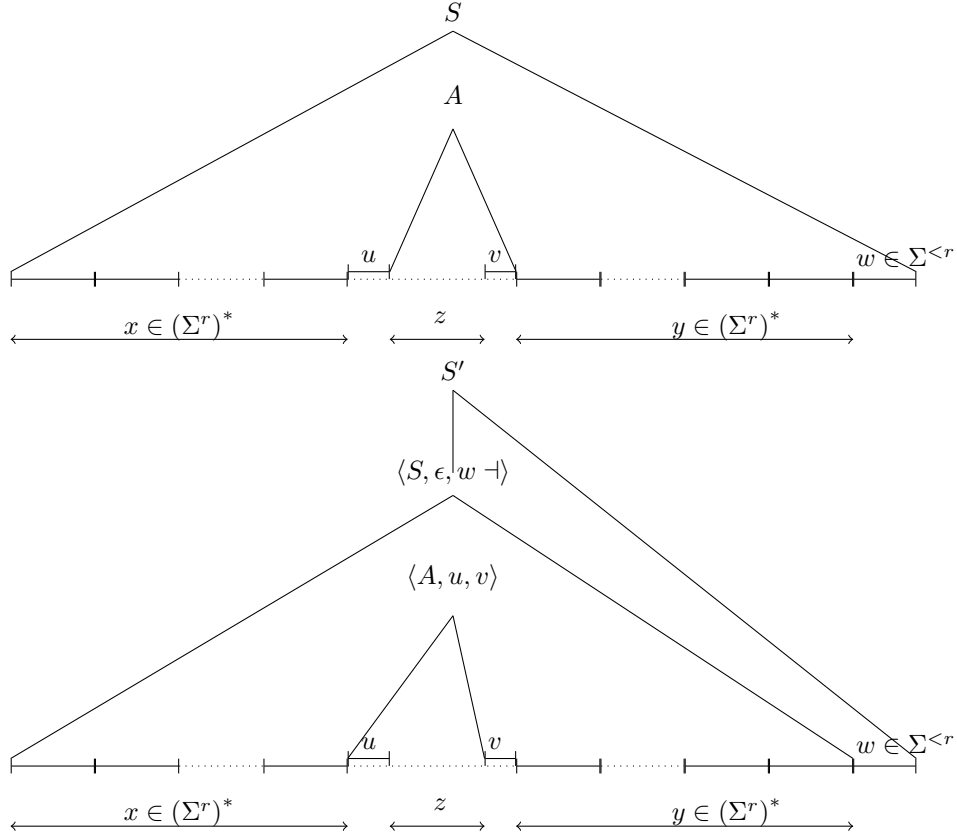
\par
Notice that the empty rules \eqref{R3} check that the presence of a letter 
  $a$ is appropriate in a specific position of the word.
Since all the terminal letters are generated by  rules  of type \eqref{R2}, every sentence of
$G'$ has a length multiple of $r$. 
At last,  the sizes of $N'$ and $P'$, respectively in \eqref{L4}  and \eqref{L5}, immediately follow from the form of the nonterminals and of the rules.
 
\par{\em Construction of $G''$.} To prove part \eqref{L6} of the thesis, we apply Lm.~\ref{lm-DGNF} 
by first modifying $G'$ into an intermediate grammar, denoted $\widehat{G} = (\Delta, \widehat{N},\widehat{P},\widehat{S})$, on the tuple alphabet $\Delta = \Delta_1 \cup \Delta_2 \dots \cup \Delta_r$, as follows. 
\par
The nonterminal alphabet $\widehat{N}$ is composed of $N'$ and of a new nonterminal $X_w$ for each rule of $G'$ of the form $S \to X w $. 
\par
The rule set $\widehat{P}$ is obtained from $P'$ by the following steps: 
\begin{enumerate} 
	\item Each rule of $G'$ of the form $S \to X w $ is replaced in $\widehat{P}$ by  two rules 
\[
S \to X X_w \; \text{ and } \; X_w \to \pi^{-1}_{|w|}(w)  
\]
 (in other words, $w$ is replaced by the corresponding tuple symbol $\pi^{-1}_{|w|}(w)$ in $\Delta_{|w|})$.
	\item
Each rule of $G'$  of the form $A \to x$,  where  $x \in \Sigma^r$,  is replaced by the rule $A \to \pi_r^{-1}(x)$.
\item Every other rule of $P'$ is in $\widehat{P}$; no other rule is in $\widehat{P}$.
\end{enumerate}
The resulting grammar $\widehat{G}$ is in CNF. The nonterminal alphabet size is $|\widehat{N|} \in O\left(|N'|+ |\Sigma|^r\right)$, which is in $O\left(|N|\cdot |\Sigma|^{2r}\right)$. 
The rule set has cardinality  $|\widehat{P}| \in O\left(|P'|\right)$.	

\par
Then, we apply Lm.~\ref{lm-DGNF} to $\widehat{G}$, obtaining a  DGNF  grammar, denoted $\widetilde{G}$, with  a number of rules 
$\left|\widetilde{P}\right| \in O\left(|\Delta|^6\cdot |\widehat{N}|^8  \right)$, which is  $O\left(|\Sigma|^{6r}\cdot |N'|^8\right)$.
Since, by~\eqref{N'},  $|N'|$ is in $O(|N|\cdot |\Sigma|^{2r})$, it immediately follows that: 
\begin{equation}\label{eq-widehatSize}
|N'|\in O\left(|\Sigma|^{22r}\cdot |N|^8\right). 
\end{equation}

\par\noindent
At last, it is immediate to transform grammar $\widetilde{G}$ back into a Q-DGNF grammar $G''$ of order $r$  over the alphabet $\Sigma$, with the same number of rules as $\widetilde{G}$.
\qed
\end{proof}

\paragraph{Digression: a useful construction of the $(m,m)$-GNF  of a CNF grammar $G$}
Incidentally, a bonus of Lm.~\ref{lm-partitionedCNF} is the direct construction of an $(m,m)$-GNF grammar, 
whose size may in general be smaller than the size produced by the standard constructions of an $(m,n)$-GNF grammar 
(e.g., the one of~\cite{DBLP:journals/ipl/Yoshinaka09}), in the special but relevant case when $m=n$. We compare 
the size of the $(m,m)$-GNF grammar obtained through the two approaches:
\[
\renewcommand{\arraystretch}{1.2}
\begin{array} {c|c}
  \text{Case $m=n$ of~\cite{DBLP:journals/ipl/Yoshinaka09}} & \text{Construction of Lm.~\ref{lm-partitionedCNF}}
	\\\hline
O\left(|\Sigma|^{2m+2}\cdot |N|^{4m+4} \right) &
O\left(|\Sigma|^{22m}\cdot |N|^{8} \right) 
\end{array}
\]
For a fixed  alphabet $\Sigma$ and a fixed value $m\ge 2$,  when considering larger and larger grammars $G$, 
the  size of our equivalent $(m,m)$-GNF grammar will eventually be smaller than the size in~\cite{DBLP:journals/ipl/Yoshinaka09}.

\subsection{Main result and proof}
We are going to prove  that, given any  terminal alphabet $\Sigma$, 
there exist a  Dyck alphabet $\Omega_{q,l}$, with $l=|\Sigma|$ neutral symbols 
and the number $q$ of brackets being polynomial in $\Sigma$, 
and a letter-to-letter homomorphism from the Dyck alphabet to $\Sigma$, such  that every CF language $L$ over $\Sigma$ has a CST characterization in terms of the Dyck language $D_{q,l}$.
We stress that the Dyck alphabet size only depends on the size of the terminal alphabet; an upper bound on the dependence is formulated later  as Corollary~\ref{corollSizeDyckAlph}.
\par 
Moreover, the regular language used in CST can be chosen to be strictly locally testable. 
\begin{theorem}\label{theorGeneralHomomCharacterization}
For every finite alphabet $\Sigma$, there exist a number $q>0$ polynomial in $|\Sigma|$ and a letter-to-letter homomorphism $\rho:\, \Omega_{q,|\Sigma|} \to \Sigma $, 
such that, for every context-free language $L\subseteq \Sigma^*$,  
there exists 
a regular language $T\subseteq \left(\Omega_{q,|\Sigma|}\right)^*$  
satisfying $L = \rho\left(D_{q,|\Sigma|} \cap T \right)$.
\end{theorem}
The proof  involves several transformations of alphabets, grammars and languages,  and relies on the constructions and lemmas presented in Sect.~\ref{subsectNormalForms}.
To improve readability, we have divided the proof into two parts.
First, we formulate in Th.~\ref{th-CST-even} a case less general than Th.~\ref{theorGeneralHomomCharacterization}, which excludes odd-length sentences from the language,
yet it already involves the essential ideas and difficulties. A step of the proof requires some arithmetic analysis, to construct  the coding that  represents the Dyck brackets, using fixed length codes ove  a smaller alphabet. In the proof, such analysis has been encapsulated  into Proposition~\ref{prop-encoding}.
\par
Second, in Sect.~\ref{SubSubSectArbitraryLength} we introduce into the Dyck language the neutral symbols that are needed for handling odd-length sentences, and we easily conclude the proof of Th.~\ref{theorGeneralHomomCharacterization}.

\subsubsection{The case of even length}\label{subsubsectCaseEvenLength}
The next theorem applies to languages of even-length words. 
Starting from the original CNF grammar $G$, we convert it to a Q-DGNF grammar of order $m$ (as in Lm.~\ref{lm-partitionedCNF}). We deal with each of the axiomatic rules $S \to X w$ at a time, by considering the subgrammar
having $X$ for axiom, which is in $(m,m)$-DGNF and defines the language $L(G,X)$. 
\par
We apply the inverse tuple homomorphism $\pi_m^{-1}$ (Def.~\ref{DefTupleAlphabet}), thus
condensing all terminal factors of length $m$ occurring in each rule into one symbol of  $\Delta_m$. The result is an almost identical grammar, here called $\widetilde{G}$, 
over the tuple alphabet $\Delta_m$ rather than $\Sigma$. 
Since $\widetilde{G}$ is  an even-DGNF and satisfies the hypothesis of  Th.~\ref{th-okhotinNostro}, 
there exists a  non-erasing CST characterization of the tuple language generated by $\widetilde{G}$. The corresponding Dyck alphabet $\Omega_k$ has however a size $k$ dependent on the size 
of $\widetilde{G}$, hence also on the size of $G$. 
\par
Now, the crucial idea comes into play.  We represent each one of the $k$ open brackets in $\Omega_k$ with an  $m$-digit integer, represented  in a base $j\ge 2$, such that only $m$ depends on the size of $G$:
we show in Proposition~\ref{prop-encoding} that, if $m$ is at least logarithmic in the size of the grammar, then there exists a suitable value of $j$, independent from the grammar, such that the open brackets are represented by codes of length $m$.

\par
To make room for such codes, we transform back  each $m$-tuple symbol of grammar $\widetilde{G}$ into a word of length $m$ (using the homomorphism $\pi_m$), 
obtaining again an $(m,m)$-GNF grammar, over a new Dyck alphabet $\Omega_q$. In such alphabet each symbol is a 4-tuple composed of:
\begin{itemize}[-]
\item
 a symbol specifying whether the bracket is open/closed;
\item
the letter of $\Sigma$ that is represented by the symbol;
\item
  the letter of $\Sigma$ that is represented by the matching closed bracket;
	\item
a digit of the code in base $j$.
\end{itemize} 
Notice that a closed bracket $\omega'\in \Omega_k$ 
is encoded as the reversal of the code that represents  the matching open bracket $\omega$; in this way,   
the string of the $m$ open brackets encoding $\omega$ is matched exactly by the  $m$ closed brackets encoding $\omega'$.
The size of the terminal alphabet of $\widetilde{G}$ does not depend on the size of $G$ and is polynomially related with the size of $\Sigma$.
\par
We then define a regular language to check whether two codes may or may not be adjacent.
Another letter-to-letter  homomorphism (denoted by $\rho$) is then used to map each 4-tuple into a letter of $\Sigma$, so that we obtain a CST characterization of $L(G,X)$ of the intended type.
\par
Next, we have  to deal with the  axiomatic rule $S \to X w$ of the Q-DGNF grammar. 
Since by hypothesis the value $|w|<m$ is an even number, it is immediate to obtain word $w$ as the homomorphic image of a Dyck language over another alphabet that  
does not depend on the size of the original grammar $G$. It is a simple matter  to combine this new part with the preceding CST characterization of $L(G,X)$, thus obtaining the CST characterization of each language $L(G,X)w$.
At last, it suffices to unite the  CST characterizations for each word $w$ and for each nonterminal $X$ such that a rule $S \to Xw$ is present in the original Q-DGNF grammar.

\begin{theorem}\label{th-CST-even}
For every finite alphabet $\Sigma$, 
there exist a number $n>0$,  polynomial in $|\Sigma|$, and a letter-to-letter homomorphism $\rho$
such that for every context-free language $L\subseteq \left(\Sigma^2\right)^*$ 
there exists 
a regular language $T\subseteq \Omega^+_n$ , 
such that $L = \rho\left(D_n \cap T \right)$, where $D_n$ is the Dyck language over the Dyck alphabet $\Omega_n$.
\end{theorem}
\begin{proof}
Let $L\subseteq \left(\Sigma^2\right)^*$ be a CF language.
Let $m\ge 2$ be an even number.
$L$ can be generated by a grammar $G = (\Sigma, N, P,S)$ in Q-DGNF of order $m$, as in Lm.~\ref{lm-partitionedCNF}.
\par
Let $w \in\Sigma^{<m}$ and let  $X \in N$ be such that $S \to Xw$ is in $P$. We  deal with the language $L(G,X)$ first.
The language $\pi_m^{-1}\left(L(G,X)\right)$ can be considered as the language generated by a grammar, 
called $\widetilde{G}$, in DGNF over the alphabet $\Delta_m$, where the rules are obtained from those of $G$ as follows:
\begin{itemize}[-]
	\item 
	ignore
all rules whose left-hand side is a nonterminal unreachable from $X$; 
	\item
	replace in the right-hand part of every other rule of $G$, every occurrence of every word $x\in\Sigma^{m}$ with the tuple symbol $\pi^{-1}_m(x)$.
\end{itemize}
The language $L(\widetilde{G})$ over the tuple alphabet can be characterized using  Th.~\ref{th-okhotinNostro}, part 2, as $h\left( D_q \cap R_X \right)$,
where $q= |\widetilde{P}|^2 +|N|\cdot |\widetilde{P}|$ is the size of the Dyck alphabet,  
$h:  \Omega_q \to \Delta_m$ is the letter-to-letter homomorphism of Th.~\ref{th-1-Okhotin},
and $D_q, R_X\subseteq {\Omega_q}^*$ are respectively a Dyck language and a regular language (which is dependent on $X$).
\par\noindent
The reason for choosing the slightly more general version in part 2 of Th.~\ref{th-okhotinNostro}, is that we later need to extend the CST characterization from 
a single language $L(G,X)$, to the  union of all languages $L(G,X)$ for $X \in N$, i.e., to  $L$.
\par\noindent
Hence, $L(G,X)=\pi_m(L(\widetilde{G}))$ and 
\begin{equation}
L(G,X)=\pi_m\left(h\left( D_q \cap R_X \right)  \right).
\label{eq:OkhotinCSTapplication}
\end{equation}
Formula \eqref{eq:OkhotinCSTapplication} is already a CST characterization for $L(G,X)$, but the value $q$
 is in $O\left(|\widetilde{P}|^2 \right)$, and $|\widetilde{P}|\in O\left(|P|\right)$; 
hence $q$ still depends on  grammar $G$.

\paragraph{Positional encoding of brackets}\label{subsubsectPositionalEncoding}
Each element $\omega \in \Omega_q$ is identified by an integer number $\iota$, with $1\le \iota \le q$.
We want to represent each of the $q$ values $\iota$ using at most $m$ digits in a base $j$:
It is enough to satisfy the inequality  $\log_j{q}\le m$. 
Denoting with $\log$ the base 2 logarithm,
this requires that $j$ satisfies $\frac{\log {q}}{\log j}\le m$, i.e., $j$ and $m$ satisfy  the inequality
\begin{equation} 
\log j \ge \frac{\log q}{m}.
\label{eqBaseInequality}
\end{equation}
If \eqref{eqBaseInequality} is satisfied, every open bracket $\omega\in \Omega_q$ can be encoded in base $j$ by a (distinct) string with $m$ digits, to be
denoted in the following as  $\left\llbracket\omega \right\rrbracket_j$. The closed parenthesis $\omega'$ matching $\omega$ has no encoding of its own, but it is
just represented by the reversal of the encoding
of $\omega$, i.e., $\left(\left\llbracket\omega \right\rrbracket_ j\right)^R$; we will see that no confusion can arise.

\par 
Although an arbitrarily large value of $j$ would satisfy \eqref{eqBaseInequality}, we prefer to  choose a value as small as possible. 
Let $p$ be the number of nonterminals of a CNF grammar defining $L$. By Lm.~\ref{lm-partitionedCNF}, in the worst case $|P| \in 
O\left(|\Sigma|^{22m}\cdot p^{8}\right)$. 
Since $q \in O(|P|^2)$, it follows that $q \in O(|\Sigma|^{44m}\cdot p^{16})$.
\par

We can abstract the expression for value $q$ as $O\left(\sigma^{m} \nu\right)$, for suitable values $\sigma=|\Sigma|^{44} , \nu= p^{16}$. 
The next proposition shows the correct numerical relation that eliminates the dependence of $j$ from $m$ and from the number of rules of the grammar.

\begin{proposition}\label{prop-encoding}
Given numbers $\sigma, m,\nu >0$, if
$m$ is in $\varOmega(\log \nu)$, then there exists $j \in O(\sigma)$ such that every 
symbol in a set of cardinality $1 <q < \sigma^{m} \nu$  can 
be represented in base $j$ by a distinct string of $m$ digits. 
\end{proposition}
\begin{subproof}
We have:
\[
\renewcommand{\arraystretch}{1.2}
\begin{array}{l}
q^{1/m}<\sigma \nu^{1/m}
\\
\log{q^{1/m}}<\log{\left(\sigma \nu^\frac{1}{m}\right)}
\\
\frac{\log q}{m}< \log{\sigma} + \frac{\log \nu}{m}, \;\text{ and }\text{if }m>\log(\nu) 
\\
\frac{\log q}{m}<  \log{\sigma} +1= \log{\left(2\sigma\right)}.
\end{array}
\]
Hence, the condition  $\log j \ge \frac{\log q}{m}$ can be satisfied, when  $m$ is in $\varOmega(\log \nu)$,
by choosing $j$ such that
$\log{j}\ge \log{\left(2\sigma\right)}$, i.e.,  $j\ge 2\sigma$. 
Thus, it suffices to choose  a suitable $j$ in $O(\sigma)$.
\qedsymbol 
\end{subproof}
From Proposition~\ref{prop-encoding} it follows that each one of the $q$ open brackets in  $\Omega_q$ can be encoded with a distinct string composed of $m$ digits in base $j\ge 2$,
with 
\begin{equation}\label{eq-jvalue}
j\in O\left(|\Sigma|^{44}\right) \text{ when } m \in  \varOmega\left(\log p^{16}\right)= \varOmega(\log p). 
 \end{equation}

\paragraph{The Dyck alphabet $\Omega_n$}
Given the values $j,m$ computed above,  
let $n=j\cdot |\Sigma|^2$, hence $n \in O\left(|\Sigma|^{46}\right)$, and define the new Dyck alphabet $\Omega_n$,  to be isomorphic to the set:
	\begin{equation}
	\label{eqDomainOmegan}
\left\{\text{`[' , `]' }\right\} \times \Sigma \times \Sigma \times \left\{ 0,\dots, j-1\right\}
	\end{equation}
Let the matching open/closed elements $\zeta, \zeta'$ in  $\Omega_n$ be:
\begin{equation}\label{eqDefMatchingPairs}
\zeta= \left\langle \text{`['}, a, b, o \right\rangle \text{ matching }\zeta'= \left\langle \text{`]'}, b, a, o \right\rangle
\end{equation}
 Note that in $\zeta$ and $\zeta'$ the second and third components are interchanged and    component $o$ is in  $0,\dots, j-1$.	
\par\noindent																														
We sum up the structure and information contained in the Dyck alphabet $\Omega_n$.  Each matching open and closed bracket, $\zeta$ and  $\zeta'$, is
represented by a 4-tuple carrying the following information:
\begin{itemize}[-]
	\item whether the element is an open or closed bracket;
	\item the letter of $\Sigma$ to which $\zeta$ will be mapped by homomorphism $\rho$;
	\item the letter of $\Sigma$ to which $\zeta'$ will be mapped by homomorphism $\rho$;
	\item a digit $i$ in the given base $j$. 
In any two matching elements $\zeta, \zeta'$, the digit $i$ is the same.
\end{itemize}
Let $D_n$ be the Dyck language over $\Omega_n$.

\paragraph{Definition and properties of homomorphism $\tau$}
We  define a new homomorphism $\tau: \Omega_q \to \Omega_n^+$
such that the image of $D_q$ by $\tau$ is  a subset of the Dyck language  $D_n$, i.e.,  $\tau(D_q) \subset D_n$. 
Such subset $\tau(D_q)$ is next obtained by means of the regular language $\tau(R_X)$, as $\tau(D_q)=D_n \cap \tau(R_X)$.
\par\noindent		
To define $\tau$, 
we first need the partial mapping, called \emph{combinator}:
\[
\otimes : (\Sigma_1)^+ \times (\Sigma_2)^+ \times (\Sigma_3)^+  \times (\Sigma_4)^+ \; \to \;\left(\Sigma_1 \times \Sigma_2 \times \Sigma_3 \times \Sigma_4\right)^+
\]
 where each $\Sigma_i$  is a finite alphabet; the mapping combines four words of identical length into one word of the same length over the alphabet of 4-tuples.
More precisely, the \emph{combinator} $\otimes$
is defined for all $\textit{l}\geq 1$, $x_i \in (\Sigma_i)^\textit{l}$ and $1\leq i \leq 4$  as:
\[
\otimes\left(x_1 , x_2 , x_3 , x_4\right) = \left\langle x_1(1),x_2(1),x_3(1),x_4(1) \right \rangle\, \dots \, \left\langle x_1(\textit{l}),x_2(\textit{l}),x_3(\textit{l}),x_4(\textit{l}) \right\rangle.
\]
For instance, let $x_1=ab, x_2=cd, x_3=ef, x_4= ca$; then $\otimes\left( x_1, x_2 , x_3 , x_4\right) = \left\langle a,c,e,c \right\rangle \, \left\langle b,d,f,a \right\rangle$.
\par\noindent
Recall now the letter-to-letter homomorphism $h: \Omega_q\to\Delta_m$, defined in the CST characterization of Eq. \eqref{eq:OkhotinCSTapplication}. 
Since $L(\widetilde{G})$ is a subset of $(\Delta_m\Delta_m)^*$, 
the image $h(\omega)$ of a  bracket $\omega\in\Omega_q$ is in $\Delta_m$. 
\par\noindent
The definition of $\tau$ is : 
\begin{gather}
\label{EqHomotau1}
\begin{array}{l}
\tau(\omega) = \otimes \left(\text{`['}^m \, ,  \pi_m \left(h(\omega) \right) \, , \left(\pi_m \left(h(\omega') \right)\right)^R
\, , \left\llbracket\omega \right\rrbracket_ j \right)
\\
\tau(\omega') =\otimes \left( \text{`]'}^m \, ,\pi_m \left(h(\omega') \right) \, , \left(\pi_m \left(h(\omega) \right)\right)^R
\, , \left(\left\llbracket\omega \right\rrbracket_j\right)^R
     \right)
\end{array}
\end{gather}
All four arguments of $\otimes$
	are words of length $m$, therefore the combinator $\otimes$ returns a word of length $m$ over the alphabet of 4-tuples.
	\par\noindent
For instance,  	if $h(\omega)= \langle a_1, \dots, a_m \rangle \in \Delta_m$,
 $h(\omega')= \langle b_m, \dots, b_1 \rangle \in \Delta_m$,
and $\left\llbracket\omega \right\rrbracket_j = o_1 o_2 \dots o_m$,
with $o_1, \dots, o_m \in \{0, \dots, j-1\},$ then  $\left(\left\llbracket\omega \right\rrbracket_ j\right)^R=o_m o_{m-1} \dots o_1$ and:
\[
	\begin{array}{llcccl}
	\tau(\omega)=& \left\langle \text{`['}, a_1, b_1, o_1 \right\rangle & 
	\left\langle \text{`['}, a_2, b_2, o_2 \right\rangle & 
	\dots & 	\dots  & \langle \text{`['}, a_m, b_m, o_m\rangle 
	\\
	\tau(\omega')=&\left\langle \text{`]'}, b_m, a_m, o_m \right\rangle &	 \dots & \dots & \left\langle\text{`]'}, b_{2}, a_{2}, o_{2}\right\rangle
& \langle \text{`]'}, b_1 , a_1, o_1 \rangle
	\end{array}
\]
An example of a complete definition of $\tau$ is given in Sec.~\ref{sect:example}, Eq.~\ref{eqExampleTau(m=2,j=2)}.
\begin{claim}\label{claimMatchingPairs}
The following  two facts hold:
\begin{enumerate}[a)]
 \item  
Let $\omega, \omega'\in \Omega_q$ be a matching pair. Then
$\tau(\omega)= \zeta_1 \dots \zeta_m$ and 
$\tau(\omega')=\zeta'_m \dots \zeta'_1$,  
where for all $i$ the pairs  $\zeta_i,\zeta'_i$ are matching in $\Omega_n$.
\item  
$\tau(D_q) \subseteq D_n$.
\end{enumerate}
\end{claim}
\begin{subproof}
Part a). The fact that $\zeta_i,\zeta'_i$  match according to formula \eqref{eqDefMatchingPairs}, follows immediately from the definition of  $\tau$.
\par\noindent
Part b). Since, for every $w \in \Omega_q^+$, $\tau(w)$ preserves
the parenthetization of $w$,  if $w \in D_q$, then $\tau(w)\in D_n$.
\qedsymbol
\end{subproof}
\par
We show that the mapping $\tau$ is one-to-one:
\begin{claim}\label{Claim1to1}
For all $ w, w' \in (\Omega_q)^+$, if $\tau(w)= \tau(w')$, then  $w=w'$.
\end{claim}
\begin{subproof}
Let $\omega_1, \omega_2 \in \Omega_q$; if  $\omega_1 \neq \omega_2$, then
$\llbracket \omega_1 \rrbracket_j \neq \llbracket \omega_2 \rrbracket_j$ by definition of $\llbracket \dots \rrbracket_j$. Therefore the inequality 
$\tau(\omega_1) \neq \tau(\omega_2)$ holds, because at least one position differs.
\qedsymbol
\end{subproof}

\paragraph{Definition and properties of the  homomorphism $\rho$ used in CST}
We now define a letter-to-letter homomorphism $\rho: \Omega_n \to \Sigma$, 
in order  to prove later that
$
\rho\left( D_n \cap \tau(R_X) \right)
$ is exactly 
$L(G,X)$. 
\par\noindent
The homomorphism $\rho$, which does not depend on the grammar $G$ but only on $\Omega_n$, is simply defined as 
the projection on the second component of each 4-tuple: 
	$
	\rho\left(\left\langle x_1,x_2,x_3,x_4 \right\rangle  \right) = x_2$\, (where $x_2 \in \Sigma$).
	
\begin{claim}\label{ClaimCommutation}
For all $ w \in (\Omega_q)^+$, the equality  $\rho(\tau(w))= \pi_m (h(w))$ holds, where $\tau$ is defined in Eq.~\eqref{EqHomotau1} and $h$ in Eq.~\eqref{eq:OkhotinCSTapplication}.
\end{claim}
\begin{subproof}
By the definitions of $\tau$ and $\rho$, for every $\chi \in \Omega_q$ the equality $\rho\left( \tau(\chi)\right)= \pi_m (h(\chi))$  holds. 
\qedsymbol
\end{subproof}

\begin{claim}\label{ClaimTau}
 $\tau^{-1}(D_n)\subseteq D_q$.
\end{claim}
\begin{subproof}
Although $\tau^{-1}$ is not defined for every word in $D_n$, mapping $\tau$ is defined so that, if a word $w\not\in D_q$,
then $\tau(w) \not\in D_n$; hence if $\tau(w)\in D_n$, then also $w \in D_q$.\qedsymbol
\end{subproof}

\paragraph{CST characterization of $L(G,X)$}
To complete this part of the proof, it is enough to prove the following identity
\begin{equation}
\rho\left( D_n \cap \tau(R_X) \right) = \pi_m \left(h \left( D_q \cap R_X \right) \right)
\end{equation}
since $L(G,X) = \pi_m \left(h \left( D_q \cap R_X \right) \right)$. 
\par\noindent
By Claim~\ref{ClaimCommutation}, $\tau(D_q) \cap \tau(R_X)= \tau(D_q \cap R_X)$;
hence, by Claim~\ref{claimMatchingPairs}, part (b),
\[
\rho\left( \tau(D_q \cap R_X) \right) \; = \;\rho\left( \tau(D_q) \cap \tau(R_X) \right) \subseteq \rho\left( D_n \cap \tau(R_X) \right).
\]
 The inclusion
\[
\pi_m \left(h \left( D_q \cap R_X \right) \right)\subseteq \rho\left( \tau(D_q \cap R_X) \right)
\]
then follows: if $z\in \pi_m (h \left( D_q \cap R_X \right)) $, then there exists a word $w \in D_q \cap R_X$
such that $\pi_m (h(w))=z$, hence
$z= \rho(\tau(w))$ by Claim~\ref{ClaimCommutation}.
Since $w\in D_q \cap R_X$, then $\tau(w) \in \tau(D_q \cap R_X)$, hence 
\[
z \in
\rho\left( \tau(D_q \cap R_X) \right)\subseteq \rho\left( D_n \cap \tau(R_X) \right).
\]
The opposite inclusion
$
\rho\left( D_n \cap \tau(R_X) \right) \subseteq \pi_m \left(h \left( D_q \cap R_X \right) \right)
$
also follows: if $z\in \rho\left( D_n \cap \tau(R_X) \right)$, then
there exists $w\in R_X$ such that  $\tau(w)\in D_n$ and $\rho(\tau(w))=z$. By Claim~\ref{ClaimTau}, if $\tau(w)\in D_n$, then also $w \in D_q$.
Since $z = \pi_m \left(h(w)\right)$ by Claim~\ref{ClaimCommutation}, it follows that $z \in  \pi_m \left(h \left( D_q \cap R_X \right) \right)$.
\par\noindent
It then follows that $L(G,X) =\rho\left( D_n \cap \tau(R_X) \right)$, where $\tau(R_X)$ is a regular language depending on the grammar $G$, while both the 
homomorphism $\rho$ and the Dyck language $D_n$ do not depend on grammar $G$, but only on $\Sigma$.
\paragraph{Extending the CST characterization}
It remains to extend the CST characterization first to $L(G,X)\cdot w$ and then to 
$\displaystyle{L = \bigcup_{X\in N, w \in \Sigma^{<m}} L(G,X)\cdot w}$
\par
First, we notice that the ``short'' word $w$, of even length, can be immediately associated with a suitable Dyck set.
Let $\sigma$ be the set $\left\{\,`[` \,\right\}\times \Sigma \times \Sigma \times \{0\}$ and let $\sigma'$ be the set $\left\{\,`]` \,\right\}\times \Sigma \times \Sigma \times \{0\}$.
Let $R_w$ be the regular language composed only of the words $\alpha \in (\sigma\sigma')^* \cap D_q$ such that $\rho(\alpha) = w$, and 
 let  $T_{X,w}= R_X \cdot R_w$. Therefore, $L(G,X)\cdot w = \rho(D_{n} \cap T_{X,w})$.
\par
The original language $L$ is the union of all $L(G,X)\cdot w$, for $X \in N, w \in \Sigma^{<m}$. 
 Set $\Omega_{n}$ was defined in Eq.~\eqref{eqDomainOmegan}, and by selecting the width $m$  and the base $j$  as in Eq.~\eqref{eq-jvalue}, 
$\Omega_{n}$ is 
large enough to encode every bracket of the Dyck alphabet $\Omega_q$ with a distinct string in $\left(\Omega_n\right)^m$. 
\par\noindent
Hence, it is immediate to define a regular language $T$ as the union of all regular languages $T_{X,w}$, for every $Xw$ such that $S \to X w \in P$. 
Therefore, $L= \rho\left(D_{n} \cap T\right)$.
\qed
\end{proof}

\begin{corollary}\label{corollSizeDyckAlphEven}
The cardinality $n$ of the Dyck alphabet of Th.~\ref{th-CST-even} is  $O(|\Sigma|^{46})$.
\end{corollary}

\paragraph{Using an SLT language}
\par
We observe that the regular language $\tau(R_X)$ in the proof of Th.~\ref{th-CST-even}
is not strictly locally testable (Def.~\ref{defk-SLT}).
Yet, it would be straightforward to modify our construction to obtain an SLT language having width in $O(\text{log } p)$: for that 
it suffices to modify homomorphism $\tau$ of Eq.~\eqref{EqHomotau1} so that the first bracket of each $\tau(\omega)$ and the last one  of $\tau(\omega')$ are  made typographically different, e.g., by using a bold font, from the remaining $m-1$ brackets
of $\tau(\omega)$ and of $\tau(\omega')$. 
\par\noindent
For instance,  if $h(\omega)= \langle a_1, \dots, a_m \rangle \in \Delta_m$ and $\left\llbracket\omega \right\rrbracket_j =o_1 o_2 \dots o_m$,
then  

\[
\tau(\omega)=\langle \text{`\textbf{(}'}, a_1, b_1, o_1\rangle \,\langle\text{`['}, a_2, b_2, o_2\rangle
	\dots \langle\text{`['}, a_m, b_m, o_m\rangle
	\]
	and 
\[\tau(\omega')=\langle \text{`]'}, b_m, a_m, o_m\rangle \dots \langle\text{`]'}, b_2 , a_2, o_2\rangle \langle\text{`\textbf{)}'}, b_1 , a_1, o_1\rangle.
\] 
Therefore, we can state: 

\begin{corollary}\label{cor-SLT}
In the CST characterization of Th.~\ref{th-CST-even}, the regular language $R$ may be assumed to be strictly locally testable. 
\end{corollary}

\subsubsection{An example}\label{sect:example}
The example illustrates the crucial part of our constructions, namely the homomorphism $\tau$ defined by Eq.~\eqref{EqHomotau1}.
Consider the  language and grammar
\[
L= \{a^{2n+4} b^{6n} \mid n\ge 0\}, \quad \{S\to aaSb^6 \mid a^4\}
\]
This grammar,  as a quotiented normal form of order 2, would be written as:
\[
S\to S_{/\varepsilon},\, S_{/\varepsilon}\to aa S_{/\varepsilon}b^6 \mid a^4
\]
We  choose the value $m=2$ for the equivalent $(m,m)$-GNF, and, in accordance, the substrings of length two occurring in the language  are mapped on the 2-tuples $\langle a,a\rangle, \langle a,b\rangle, \langle b,b\rangle$, shortened as $\langle aa\rangle$, etc.
\par
\begin{figure}[h]
\begin{center}
\scalebox{0.7}{
\begin{tikzpicture}[auto, level 1/.style={sibling distance=60mm},
    level 2/.style={sibling distance=24mm},
    level 3/.style={sibling distance=12mm},xscale=0.9]
\node{$1:S$}
 child { node{$<aa>$}}
 child { 
  node{$1:S$}
		child { node{$<aa>$}}
		child { 
		node {$2:S$}
			child{ node {$<aa>$}} child{ node { $<aa>$}} 
		}
		child { 
		node {$3:B$}
		child{ node {$<bb>$}} child{ node {$<bb>$}} 
				}
			child { node {$<bb>$}}	%
			}
		child { 
		node {$3:B$}
		child{ node {$<bb>$}} child{ node {$<bb>$}} 
		}
  child { node {$<bb>$}}
  ;    
\end{tikzpicture}
}
\end{center}
\caption{Syntax tree of the sentence $a^8 b^{12} \in L$ after its transformation to $ \langle aa\rangle^4 \langle bb\rangle^6 \in L(G'')$.}
\label{fig:exampleTrees}
\end{figure}
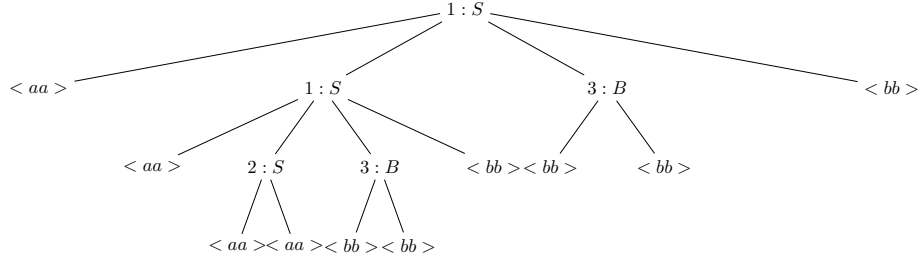

\noindent
The following grammar in DGNF, though  constructed by hand, takes the place of grammar  $G''$ of Lm.~\ref{lm-partitionedCNF}:
\begin{equation}
G'' \;= \; \Big\{
1: S \to \langle aa\rangle \, S  \,B\,  \langle bb\rangle , \;
 2: S\to \langle aa\rangle \,\langle aa\rangle , \;
 3:B \to \langle bb\rangle \, \langle bb\rangle
\Big\}.
\label{eqExampleGrammOver2ples}
\end{equation}
The sentence $a^8 b^{12} \in L$ becomes $ \langle aa\rangle^4 \langle bb\rangle^6 \in L(G'')$, with the syntax tree in Fig.~\ref{fig:exampleTrees}. 
\par\noindent
For Okhotin Th. 1~\cite{Okhotin2012}, this sentence is the image by homomorphism $h$ of the following sequence 
\begin{equation}
\gamma=
(^{-}_{1} \quad  (^{1}_{1} \quad  (^{1}_{2}\quad
  )^{1}_{2} \quad  (^{1}_{3}\quad   )^{1}_{3} \quad  )^{1}_{1} \quad  (^{1}_{3} \quad   )^{1}_{3}\quad )^-_{1}
	\label{eqExampleOkhotinHomo}
\end{equation}
 of labeled parentheses, where the numbers identify the rules and the dash (as in~\cite{Okhotin2012}) means the root of the tree. The homomorphism is specified by the table:
\setlength{\arraycolsep}{0.8cm}
\begin{equation}
\begin{array}{c|c | c|c}
\omega & \omega' & h(\omega) & h(\omega')\\\hline
(^{-}_{1} & )^-_{1} &  \langle aa \rangle &  \langle bb \rangle
\\
(^{1}_{1} & )^1_{1} & \langle aa \rangle  &   \langle bb \rangle
\\ (^{1}_{2} & )^1_{2} &  \langle aa \rangle&  \langle aa \rangle
\\
(^{1}_{3} & )^1_{3} &   \langle bb \rangle &   \langle bb \rangle
\end{array}\label{eq-ex-h-hom}
\end{equation}
\par\noindent
Applying Proposition~\ref{prop-encoding}, we choose to represent each such labeled parenthesis with a sequence of $m=2$  digits, on the basis $j=2$. 
Therefore the homomorphism $\tau$ resulting from  Eq.~\eqref{EqHomotau1} defines the following Dyck alphabet:
\setlength{\arraycolsep}{0.8cm}
\begin{equation}
\begin{array}{c|c | c|c}
\omega & \omega' & \tau(\omega) & \tau(\omega')\\\hline
(^{-}_{1} & )^-_{1} &  [_{a,b,0} \; [_{a,b,0} &  ]_{b,a,0} \; ]_{b,a,0}
\\
(^{1}_{1} & )^1_{1} &  [_{a,b,0} \; [_{a,b,1} &  ]_{b,a,1} \; ]_{b,a,0}
\\ (^{1}_{2} & )^1_{2} &  [_{a,a,1} \; [_{a,a,0} &  ]_{a,a,0} \; ]_{a,a,1}
\\
(^{1}_{3} & )^1_{3} &  [_{b,b,1} \; [_{b,b,1} &  ]_{b,b,1} \; ]_{b,b,1}
\end{array}	
\label{eqExampleTau(m=2,j=2)}
\end{equation}
To finish, we show the value of  $\tau\left(\pi_m (h(\gamma))\right)$: 
\setlength{\arraycolsep}{0.7cm}
\begin{equation}
\begin{array}{l}
  \overbrace{[_{a,b,0}\; [_{a,b,0}}^{(^{-}_{1}}\;
	\overbrace{[_{a,b,0} \;  [_{a,b,1}}^{(^{1}_{1}}
 \overbrace{[_{a,a,1} \;  [_{a,a,0}}^{(^{1}_{2}} \;
  \overbrace{]_{a,a,0} \;  ]_{a,a,1}}^{)^{1}_{2}} \;
	\overbrace{[_{b,b,1} \;  [_{b,b,1}}^{(^{1}_{3}}
  \overbrace{]_{b,b,1} \;  ]_{b,b,1}}^{)^{1}_{3}}
	\\
	\overbrace{]_{b,a,1} \;  ]_{b,a,0}}^{)^{1}_{1}}
		\overbrace{[_{b,b,1} \;  [_{b,b,1}}^{(^{1}_{3}}
  \overbrace{]_{b,b,1} \;  ]_{b,b,1}}^{)^{1}_{3}}
	  \overbrace{]_{b,a,0}\; ]_{b,a,0}}^{)^-_{1}}
		\end{array}
		\label{exampleTauPiH}
\end{equation}
Notice that the 2-SLT language of the classical CST (applied to language $L$) is now replaced by an SLT language of higher width.

\subsection{Homomorphic characterization for languages of words of arbitrary length}\label{SubSubSectArbitraryLength}
At last, we  drop the restriction  to even-length sentences, thus  obtaining the homomorphic characterization stated in Th.~\ref{theorGeneralHomomCharacterization} that holds for any  language. 
\par
As defined in Def.~\ref{defAlphabetDyckWithNeutral}, 
let $\Omega_{q,l}$ be an alphabet with $q$ pairs of brackets and $l\ge 1$ neutral symbols, and $D_{q,l}$ be the corresponding Dyck language with neutral symbols.
\par 
In our treatment, there are exactly $l = |\Sigma|$  neutral symbols that we represent as 4-tuples of the form  $\langle -, a, a, 0\rangle$ where ``$-$'' is a new symbol. Then 
$\Omega_{q,l}= \Omega_q \cup \left\{\langle -, a, a, 0\rangle \mid a \in \Sigma\right\}$.
\par
Suppose that $L$ has also words of odd length. We still can apply Lm.~\ref{lm-partitionedCNF} to convert its grammar into a Q-DGNF grammar $G$ of (even) order $m$. 
Let $x \in L$ have odd length. Since its length is not multiple of $m$, $x$ is derived from the axiom   using a rule of $G$ of the form  
$S \to X w$, for some $X \in N$, $|w|<m$ -- we remind that $L(G,X)$ generates a language of words whose length is a multiple of $m$. 
The word $w$ can be factored as $w = w' a$, with $w'\in(\Sigma^2)^*$, $a\in\Sigma$.
Therefore, the same construction of the proof of Th.~\ref{th-CST-even} may be applied,  by finding a CST characterization
for $L(G,X)$ and then extending it  also to $L(G,X)\cdot w'$.
Hence, there exists a word $s$ over the Dyck alphabet $\Omega_{q}$
such that $w' = \rho(s)$. Just concatenate $\langle -, a, a, 0\rangle$ to the right of $s$,
and extend the definition of $\rho$ by setting $\rho\left(\langle -, a,a,0\rangle\right)=a$ for all $a\in \Sigma$. 
\par\noindent
This completes the proof of Th.~\ref{theorGeneralHomomCharacterization}.
\qed
\begin{example}\label{exHomDyckNeutral}
To illustrate the case of odd length sentences, we modify the language and grammar in the example of Sect. \ref{sect:example} as follows.
\begin{equation}
\begin{array}{l}
L= \{a^{2n+4} b^{6n} \mid n\ge 0\}\cdot c
\\
\text{A grammar for $L$ (axiom $A$): } \; A\to Sc \, , \, S\to aaSb^6 \mid a^4
\\
\end{array}
\end{equation}
Then the quotiented normal form of order 2 is the grammar 
\[
A\to A_{/c}\,c\, ,\, A_{/c}\to aa S_{/\varepsilon}b^6 \mid a^4
\] 
Few changes are needed with respect to the previous example. 
The Dyck alphabet and  homomorphism $\tau$ of Eq. \eqref{eqExampleTau(m=2,j=2)} are extended with the neutral symbol $\langle -, c, c, 0\rangle$;
e.g., the Dyck sentence of Eq.~\eqref{exampleTauPiH} needs to be concatenated with $\langle -, c, c, 0\rangle$.
The
homomorphism  $\rho$ in the statement of Th.~\ref{theorGeneralHomomCharacterization} is defined by extending $\rho$ 
of Th.~\ref{th-CST-even} with 
$\rho\left(\langle -, c, c, 0\rangle \right) = c$.
\end{example}

\section{Complexity of Dyck alphabet and relation with Medvedev theorem}\label{SectHomCharBasedOnMedvedev}
We have already given in Corollary~\ref{corollSizeDyckAlphEven} the  size of the Dyck alphabet used by Th.~\ref{th-CST-even}. Since the number of  neutral symbols introduced in Sect. \ref{SubSubSectArbitraryLength}  only linearly depends on  $|\Sigma|$, we have:
\begin{corollary}\label{corollSizeDyckAlph}
The cardinality of the Dyck alphabet $\Omega_{q, |\Sigma|}$ of Th.~\ref{theorGeneralHomomCharacterization} is  $O(|\Sigma|^{46})$.
\end{corollary}
The value $q$ is thus polynomial in the cardinality of the alphabet $\Sigma$. 
The current bound is related to our constructions of grammars in the generalized DGNF of some order $m$. 
A trivial lower bound is $\varOmega(|\Sigma|)$, but it is open whether one can always use a significantly smaller alphabet than the one computed above. 
\vspace{5mm}
\par
On the other hand,  it is easy to see that in the case of some linear grammars, the  bound of Corollary~\ref{corollSizeDyckAlph} is largely overestimated.
In particular, suppose that a grammar is both linear and in DGNF, i.e., its rules are in 
$
\left(N \times \Sigma (N\cup\{\epsilon\}) \Sigma \right) \cup \left(N \times \Sigma \right)
$.
Such grammars  generate only a subset of the linear languages, but they are still an interesting case.
\par
We now  proceed to characterize the languages generated by linear grammars in DGNF through a CST using  a different approach, based on Medvedev's homomorphic characterization of regular languages. 
In~\cite{DBLP:journals/ijfcs/Crespi-ReghizziP12} we  extended the historical Medvedev theorem~\cite{Medvedev1964,Eilenberg74}, 
which states that every regular language $R$ can be represented as  a letter-to-letter homomorphism of a 2-SLT language  over a larger alphabet. Moving beyond width two, we proved the following relation between
the alphabet sizes,  the complexity of language $R$ (measured by the number of states of its NFA),  and the SLT width parameter.

\begin{theorem}\label{theoMedvedevEsteso}\emph{\cite{DBLP:journals/ijfcs/Crespi-ReghizziP12}}
Given a finite alphabet $\Delta$, if a regular language $R \subseteq \Delta^*$ is accepted by an NFA with $|Q|$ states, then 
there exist a letter-to-letter homomorphism  $f$ and an $s$-SLT language $T$ over an alphabet  $\Lambda$ of size $2 |\Delta|$, such that
$R =  f(T)$, with the width parameter $s\in \Theta(\log{ |Q|})$.
\end{theorem} 
Our work~\cite{DBLP:journals/ijfcs/Crespi-ReghizziP12} also exhibits a language $R\subseteq\Delta^*$ such that, for any SLT language $T$ over an alphabet  of size $<2 |\Delta|$, a letter-to-letter homomorphism $f$ satisfying $R =  f(T)$ does not exist.
\par
Next, we apply Th.~\ref{theoMedvedevEsteso} to languages generated by  linear grammars in DGNF.

\begin{proposition}\label{propo:CSTforSimpleSymmLinGram}
Let $L=L(G)\subseteq \Sigma^+$, where $G=(\Sigma, N, P, S)$ is a linear grammar in DGNF. 
Then there exist a Dyck alphabet (with neutral symbols) $\Omega_{n,\l}$, a letter-to-letter homomorphism $g: \Omega_{n,\l} \to \Sigma$ and an SLT language $U$ over  $\Omega_{n,l}$ such that: 
\begin{enumerate}
	\item  $ L=g\left(D_{n,\l} \cap U \right)$,
	\item $n = 2\cdot |\Sigma|^2$ and $l = |\Sigma|$,
	\item  $U$ is an $s$-SLT language with $s \in \log(|N|)$.
\end{enumerate}
\end{proposition} 

\begin{proof}
\newcommand{\Op} {{\mathbf o}}
\newcommand{\Cl} {{\mathbf c}}
\par
For brevity, we prove the case when $L$ has only words of even length, hence neutral symbols are not needed in the Dyck alphabet. The extension to the general case is immediate.
\par\noindent
Let $\Sigma_1, \Sigma_2$ be alphabets; for all pairs $\langle a,b\rangle \in \Sigma_1 \times \Sigma_2$, 
let 
$|_1,\, |_2$ be the projections respectively on the first and the second component, i.e., $\langle a,b\rangle|_1 = a,\, 
\langle a,b\rangle|_2 = b$.
\par\noindent
Let $\Delta = \Sigma \times \Sigma$.
From the structure of  linear grammars in DGNF, it is obvious that there exists a regular language $W$ over the alphabet $\Delta$,  such that:
\begin{equation}
L= \left\{w|_1 \cdot w|_2^R \,\mid \,w\in W \right\} 
\label{eq:simpleEvenLinLanguage}
\end{equation} 
where $w|_2^R$ (equivalent to $w^R|_2$) is the mirror image of the projection $ w|_2$. 
Moreover, $W$ can be easily defined by means of an NFA  having $
|N| + 1$ states.
\par\noindent
By Th.~\ref{theoMedvedevEsteso},  there exist an alphabet $\Lambda$  of size $
n=2\cdot|\Delta|= 2\cdot|\Sigma|^2$,  a homomorphism $f: \Lambda\to\Delta$, a value $s\in \Theta(\log{ |N|})$, and an $s$-SLT language $T\subseteq \Lambda^*$ such that
$W=f(T)$.
\par\noindent
Let  $\Op \circledast \Lambda = \{\,\Op \,\} \times \Lambda$ and $\Cl \circledast \Lambda =   \{\,\Cl\,\} \times \Lambda$  be
two sets of opening and closing brackets,  stipulating that, for every $\lambda \in \Lambda$, 
bracket $\langle \Op , \lambda\rangle$ matches  bracket $\langle \Cl, \lambda\rangle$. 
Thus $\Omega_n=\left(\Op \circledast \Lambda\right) \cup \left(\Cl \circledast \Lambda \right)$ is a Dyck alphabet  
and we denote the corresponding Dyck language by $D_n$.
\par\noindent
We also define  $\Op \circledast \lambda_1 \dots \lambda_m \in \Lambda^+$, for every 
$\lambda_1 \dots \lambda_m \in \Lambda^+$, $m \ge 1$, as 
$ \langle \Op , \lambda_1\rangle   \dots \langle \Op , \lambda_n\rangle$.
The notation is further  extended to a language as usual, e.g., $\Op \circledast X$ for a language $X\subseteq \Lambda^+$ is the set of words 
$\{\Op \circledast x \mid x \in X\}$. 
The similar notations $\Cl \circledast x$ and $\Cl \circledast X$ have the obvious meaning,
 e.g., $\Cl \circledast (\lambda_1  \dots \lambda_m)$, $m \ge 1$,
is the word $\langle \Cl , \lambda_1\rangle  \dots \langle \Cl , \lambda_n\rangle$.

\par\noindent
We define the regular language $U$ over the  alphabet $\Omega_n$ as:
\begin{equation*}
U = \left( \Op \circledast T \right) \cdot \left( \Cl \circledast \Lambda^+ \right)
\end{equation*}
Since $T$ is  $s$-SLT, it is obvious that also $\Op \circledast T$ and  $U$ are $s$-SLT.
\par\noindent
It is easy to see that, for all $t= \lambda_1 \dots \lambda_m \in T$, 
the set  $\left(\Op \circledast t \right) \cdot \left(\Cl \circledast \Lambda^+ \right) \; \cap D_n\, \subseteq U$  
is a singleton including the word $\left(\Op \circledast t \right) u\in D_n$, where 
$u|_1\, = \Cl^{|t|}$ and $u|_2 = t^R$, i.e., $u= \Cl \circledast (\lambda_n  \dots \lambda_1)$.
 We can then write $u$ as the mirror image $(\Cl \circledast t)^R$ of $\Cl \circledast t$.  
\par\noindent
Denote with $U(t)$ the word $(\Op \circledast t)\cdot (\Cl \circledast t)^R$ for every $t \in T$: we have that 
$U \cap D_n = \bigcup_{t \in T} U(t)$. 
\par\noindent
Define the homomorphism $g: (\Cl \circledast \Lambda) \cup (\Op \circledast \Lambda)\to \Sigma$ as:

\[
\begin{displaystyle}
\left.
  \begin{cases}
     g(z) = \left( f(z|_2) \right)|_1 , & \text{if } 
z \in \Op \circledast \Lambda 
\\
g(z) = \left(f(z|_2)\right)|_2 , & \text{if }  z \in \Cl \circledast \Lambda 
  \end{cases}
  \right. 
\end{displaystyle}
\]
 where $f$ is the homomorphism of Th. \ref{theoMedvedevEsteso} defined above.
This definition is exemplified by 
\[
\begin{displaystyle}
\left.
  \begin{cases}
  g\left(\langle \Op ,\lambda\rangle \right) = f(\lambda)|_1  &
\\
g\left(\langle \Cl,\lambda\rangle \right) = f(\lambda)|_2&
  \end{cases}
  \right. 
\end{displaystyle}
\]
By definition, for every $t \in T$, it holds:
\[
\begin{displaystyle}
  \begin{cases}
	g\left( U(t)\right ) 	&=g(\Op \circledast t)\cdot g(\Cl \circledast t)^R
\\
										&= \left(f\left((\Op \circledast t)|_2\right) \right)|_2 \cdot  \big( f((\Cl \circledast t)|_2)|_1  \big)^R 
\\
										&= f(t)|_1 \cdot  \left( f(t)|_2 \right)^R.
	\end{cases}
\end{displaystyle}
\]
We now show that $L= g(D_n \cap U)$.
If  $x \in L$, then by Eq.~\eqref{eq:simpleEvenLinLanguage} $x = w|_1\,w|_2^R$ for some $w \in W$. 
Since $W= f(T)$, there is $t \in T$ such that $w=f(t)$ and, by definition of $U$, it holds $U(t) \in U$.
Hence, $g\left(U(t)\right)= f(t)|_1 \left(f(t)|_2\right)^R = w|_1 w|_2^R = x$.  
\par
The converse proof is similar. 
If $x \in g(D \cap U)$ then there exists $t \in T$ such that $x=g(U(t))$.  
Let $w = f(t)$, hence, $w \in W$. 
By definition of $g$, 
$x=g(U(t))= f(t)|_1 \left(f(t)|_2\right)^R = w|_1 w|_2^R$, hence $x \in L$.
\qed
\end{proof}


 \section{Conclusion}\label{SectConclusion}
 The main contribution of this paper is  the  homomorphic characterization of context-free languages using a grammar-independent Dyck alphabet and a  non-erasing homomorphism. It substantially departs from previous characterizations which either used a grammar-dependent alphabet, or had to erase an unbounded number of brackets.
 \par
 Our result says that, given a terminal alphabet, any  language over  the same alphabet  can be homomorphically characterized using the \emph{same} Dyck language and the \emph{same} homomorphism together with a language-specific  regular  language. In other terms,  the idiosyncratic properties of each context-free language are completely represented in the words of the regular language, which moreover have the same length as the original sentences. In this way, for each source alphabet size, a one-to-one correspondence between  context-fre grammars and regular (more precisely strictly locally testable) languages is established. In accordance with the trade-off between the complexity of the language, the Dyck alphabet size and the regular language complexity (Proposition~\ref{propos:TradeOff}), the more complex the source language, the higher is the width of the sliding window used by the regular language. 
 We hope that further studies of such correspondence between the two fundamental context-free and regular language families,  may lead to new insights. 
 \par
 A  technical question is open to further investigation. The Dyck alphabet size that we have proved to be sufficient  is a rather high power of the source alphabet size. It may be possible to obtain substantial size reductions, for the general case and, more likely, for some subfamilies of context-free languages, as we have shown for the  linear languages in double Greibach normal form.

\paragraph{Acknowledgment} We gratefully thank the anonymous reviewers for their careful and valuable suggestions.
\bibliographystyle{elsarticle-num}
\section*{\refname}
\bibliography{automatabib}

\end{document}